\documentclass[11pt,preprint]{aastex}
\usepackage{graphics}
\usepackage{amssymb}
\usepackage{amsmath}
\usepackage[dvips]{color}
\newcommand{\kms}{{km~s$^{-1}$}}

\shorttitle{\ion{Mg}{2} BAL quasars}
\shortauthors{Zhang et al. 2009}

\begin{document}

\title{Low-$z$ \ion{Mg}{2} Broad Absorption-Line Quasars from the Sloan Digital Sky Survey}
\author{
Shaohua~Zhang\altaffilmark{1,2},
Ting-Gui~Wang\altaffilmark{1,2},
Huiyuan~Wang\altaffilmark{1,2},
Hongyan~Zhou\altaffilmark{1,2},
Xiao-Bo~Dong\altaffilmark{1,2},
Jian-Guo~Wang\altaffilmark{3,4,5}
}

\altaffiltext{1}{Key Laboratory for Research in
Galaxies and Cosmology, The University of Sciences and Technology of
China, Chinese Academy of Sciences, Hefei, Anhui 230026, China;
~zsh,\,whywang@mail.ustc.edu.cn, twang@ustc.edu.cn}
\altaffiltext{2}{Department of Astronomy, University of Science and Technology of China,
Hefei, Anhui 230026, China}
\altaffiltext{3}{National Astronomical Observatories/Yunnan
Observatory, Chinese Academy of Sciences, P.O. Box 110, Kunming,
Yunnan 650011, China}
\altaffiltext{4}{Graduate School of the
Chinese Academy of Sciences, 19A Yuquan Road, P.O. Box 3908, Beijing
100039, China}
\altaffiltext{5}{Laboratory for the Structure and Evolution of Celestial Bodies,
Chinese Academy of Sciences, PO Box 110, 650011 Kunming, China}

\begin{abstract}
We present a sample of 68 low-$z$ \ion{Mg}{2} low-ionization broad
absorption-line (loBAL) quasars. The sample is uniformly selected
from the Sloan Digital Sky Survey Data Release 5 according to the
following criteria: (1) redshift $0.4<z\leq0.8$, (2) median spectral
$S/N>7~$pixel$^{-1}$, and (3) \ion{Mg}{2} absorption-line width
$\Delta v_{c} \geq 1600~$\kms. The last criterion is a trade-off between
the completeness and consistency with respect to the canonical
definition of BAL quasars that have the `balnicity index' $BI>0$ in
\ion{C}{4} BAL. We adopted such a criterion to ensure that $\sim
90\%$ of our sample are classical BAL quasars and the completeness
is $\sim 80\%$, based on extensive tests using high-$z$ quasar samples
with measurements of both \ion{C}{4} and \ion{Mg}{2} BALs.  We found
(1) \ion{Mg}{2} BAL is more frequently detected in quasars with
narrower H$\beta$ emission-line, weaker [\ion{O}{3}] emission-line,
stronger optical \ion{Fe}{2} multiplets and higher luminosity. In
term of fundamental physical parameters of a black hole accretion
system, loBAL fraction is significantly higher in quasars with a
higher Eddington ratio than those with a lower Eddington ratio.
The fraction is not dependent on the black hole mass in the range
concerned. The overall fraction distribution is broad, suggesting a
large range of covering factor of the absorption material. (2)
[\ion{O}{3}]-weak loBAL quasars averagely show undetected
[\ion{Ne}{5}] emission line and a very small line ratio of
[\ion{Ne}{5}] to [\ion{O}{3}]. However, the line ratio in non-BAL
quasars, which is much larger than that in [\ion{O}{3}]-weak loBAL
quasars, is independent of the strength of the [\ion{O}{3}] line.
(3) loBAL and non-loBAL quasars have similar colors in near-infrared
to optical band but different colors in ultraviolet.
(4) Quasars with \ion{Mg}{2} absorption
lines of intermediate width are indistinguishable from the non-loBAL
quasars in optical emission line properties but their colors are similar
to loBAL quasars, redder than non-BAL quasars.
We also discuss the implication of these results.
\end{abstract}


\section{Introduction}

About 15\% of quasars show broad absorption lines (BALs) of high
ionization ions such as \ion{N}{5}, \ion{C}{4}, \ion{Si}{4},
Ly$\alpha$, \ion{O}{6}, up to a velocity of $v\sim 0.1~c$. BALs are
detected occasionally (another $\sim$15\%) also in low ionization
species such as \ion{Mg}{2}, \ion{Al}{3}. Two very different
scenarios have been proposed to explain the BAL phenomenon. The
first scenario, namely `unification model', suggests that BAL and
non-BAL quasars are physically the same, and attributes their
different appearance solely to different line of sight. According to
the unification model, every quasar has a BAL region (BALR) with a
covering factor of 10\%-20\%, and our line of sight passes through
BALR only in BAL quasars, plausibly at low inclination angles
(Tolea et al. 2002; Hewett \& Foltz 2003; Reichard et al. 2003b;
Trump et al. 2006; Gibson et al. 2009, hereafter G09). The second,
so called evolutionary scenario, suggests that BAL quasars are in
the early stage of quasar evolution with a gas/dust richer nuclear
environment(Sanders et al. 1988; Hamann \& Ferland 1993; Voit et al.
1993; Egami et al. 1996; Becker et al. 2000; Trump et al. 2006).

On the one hand, there are many pieces of observational evidence for the
unification of BAL and non-BAL quasars, including the similarity of
emission line spectrum between the two classes of quasars (Weymann
et al. 1991), a small covering factor of BALR inferred from emission
line profiles (Korista et al. 1993), spectropolarimetric observations
of BAL quasars (e.g., Goodrich \& Miller 1995; Cohen et al. 1995;
Hines \& Wills 1995; Ogle et al. 1999; Schmidt \& Hines 1999),
and the great similarity of the  spectral energy
distribution (SED) between BAL and non-BAL quasars in the infrared
to millimeter waveband (e.g., Willott et al. 2003;
Gallagher et al. 2007). The notoriously weak X-ray emission from BAL
quasars is often ascribed to strong absorption in the BAL direction,
which is also supported by X-ray spectroscopy (e.g., Green et al.
1995; Brinkmann et al. 1999; Wang et al. 1999; Brandt et al. 2000;
Gallagher et al. 2002, 2006; Fan et al. 2009). On the other hand,
there are observations that cannot be understood in the simple unification
scenario. First, radio morphology and radio variability study showed
that BAL quasars are not observed at any preferred direction with
respect to the radio axis (Jiang \& Wang 2003; Brotherton et al.
2006; Zhou et al. 2006b; Ghosh \& Punsly 2007; Wang et al. 2008a).
Second, Boroson (2002) found that BAL quasars on average have higher
Eddington ratios than non-BAL quasars in a small sample of BAL QSOs.
A similar conclusion has been
reached by Ganguly et al. (2007). It has also been suggested that
BAL quasars are redder and more luminous than other quasars
(Reichard et al. 2003b, Trump et al. 2006, cf., G09).
The latter results indicate that BAL and non-BAL quasars can be unified,
but the covering factor of BALR depends on their nuclear parameters.
A wide range of covering factor of BALR has also been implied by
comparison of the optical polarization between BAL and non-BAL
quasars (Wang et al. 2005).

Significant differences between low-ionization BAL (loBAL) and 
high-ionization BAL
(HiBAL)/non-BAL quasars are also seen in dust extinction and
far-infrared emission. loBAL quasars show a redder spectrum than
HiBAL and non-BAL quasars on average, consistent with a reddening of
$E(B-V)\sim$0.1 for a SMC-like dust extinction curve (Weymann et al.
1991; Richards et al. 2003). Dai et al. (2008) showed that BAL
fraction among Two Micron All Sky Survey(2MASS) selected quasars 
are as high as $\sim$ 44\%
(cf. Ganguly \& Brotherton 2008). Surprisingly, when going down to
a low flux limit in near infrared, Maddox \& Hewett (2008) found a
similar 30\% fraction of BAL quasars. This indicates that BAL
quasars on average are heavily reddened and thus many red BAL quasars
have been overlooked in optical spectroscopic surveys.
This can be interpreted physically in two very different scenarios:
either BAL quasars
are a distinct population with the nuclei preferring a gas and dust
rich environment; or dust is preferably distributed in the outflow
direction as suggested by the dusty disk wind models for BALR
(Konigl \& Kartje 1994), and overall covering factor is 30\%.
Isotropic properties, such as far-infrared emission, are of great
importance to distinguish between the two. Boroson \& Meyers (1992)
found that the fraction of loBAL quasars in a small far infrared
selected sample is much higher than that in optically selected
samples. They also found that these quasars show weak [\ion{O}{3}]
and strong optical \ion{Fe}{2} emission lines. A large sample of
low-$z$ loBAL quasars are needed to confirm these findings.

Low-$z$ BAL quasars are of great interest also because a number of
important spectral diagnostics can be accessed via the ground optical
spectroscopic observations, such as narrow emission-lines (NELs),
Balmer and \ion{Fe}{2} broad emission lines (BELs). We can also inspect the
properties of the host galaxies much easier at low-$z$. However,
previous studies mainly focused on high-$z$ BAL or HiBAL quasars due
to the rarity of loBAL quasars. With the advent of large area
spectroscopic surveys, such as the Sloan Digital Sky Survey (SDSS;
York et al. 2000), it is possible to perform a systematic study of
low-$z$ BAL quasars based on a large sample.

In this paper, we present a sample of 68  \ion{Mg}{2} BAL
quasars at  $0.4<z\leq0.8$ uniformly selected from the quasar 
sample of SDSS Data Release 5 (Schneider et al. 2007), and make a comparison study with
non-BAL quasars. This paper is oraganized as follows. We analyze the
SDSS spectra and compile the sample in \S2. Various definition
criteria have been adopted in previous studies, and different
criteria result in different fractions of BAL quasars. We have made
extensive tests for different criteria in order to obtain a more
objective definition of loBAL quasars. The tests are described in
detail in this section, together with the definition criterion that is
used in our sample. In \S3, we compare the properties
between loBAL and non-BAL quasars. The implication of our findings
is discussed in \S4. Throughout  this paper, we assume a
$\Lambda$-dominated cosmology with $H_0=72$\kms~Mpc$^{-1}$,
$\Omega_M=0.28$, and $\Omega_\Lambda=0.72$.

\section{Sample Compilation and Data Analysis}

We start from the SDSS DR5 quasar catalog (Schneider et al. 2007),
and select 34037 quasars with redshifts of $0.4<z<1.97$ and a median
signal-to-noise ratio of $S/N>7$pixel$^{-1}$ as the parent sample of
\ion{Mg}{2} BAL quasars. The $S/N$ ratio threshold is introduced to
ensure reliable measurements of continua and absorption and emission
lines. The redshift cutoffs are so chosen that \ion{Mg}{2} falls
in the wavelength coverage of the SDSS spectrograph ($\sim
3800-9200$ \AA). We split the parent sample into `low-$z$ ($N=7229$)',
`moderate ($N=16621$)' and `high-$z$ ($N=10187$) samples with
$0.4<z\leqslant0.8$, $0.8< z\leqslant 1.53$, and $1.53<z<1.97$,
respectively. We will focus on the low-$z$ sample and cull our low-$z$
\ion{Mg}{2} BAL quasars according to an objective criterion
obtained from extensive tests using the high-$z$ sample with both
\ion{Mg}{2} and \ion{C}{4} measured. The SDSS spectra are corrected
for the Galactic extinction using the extinction map of Schlegel et
al. (1998) and the reddening curve of Fitzpatrick (1999), and
transformed into the rest frame using the redshifts in Schneider et
al. (2007) before further analysis.

\subsection{Fitting of \ion{Mg}{2} Spectral Regime}
We fit the spectrum of \ion{Mg}{2} regime in the following three
steps.
\begin{description}
\item[(1) Determination of continuum and UV \ion{Fe}{2} multiplets.] We
fit the SDSS spectra in the rest-frame wavelength range of
$2100-3100$\AA~with a combination of  a single power-law and an \ion{Fe}{2} template. The
initial values of the normalization and power-law slope are
estimated from continuum windows ([1790, 1830]\AA,
[2225, 2250]\AA~ and [4020, 4050]\AA)
that are not seriously contaminated by
emission-lines (e.g., Forster et al. 2001). We adopt the \ion{Fe}{2}
template derived by Vestergaard \& Wilkes (2001) and convolve it
with a Gaussian kernel in velocity space to match the width of
\ion{Fe}{2} multiplets in the observed spectra. Prominent
emission-lines other than \ion{Fe}{2} are masked during the fit.
\item[(2) Modeling \ion{Mg}{2} emission line.] After subtracting the modeled
power-law continuum and \ion{Fe}{2} multiplets, we fit the residual
spectrum around \ion{Mg}{2} with one or two Gaussians.
We find that two Gaussians are sufficient to reproduce \ion{Mg}{2}
line profile in most cases, and adding more component does not significantly
improve the fit.
The weights of \ion{Mg}{2} blue wing are reduced to half of that of
the red wing in order to lessen the effect of potential \ion{Mg}{2}
absorption troughs.
\item[(3) Remodeling of the spectra.] We mask the spectral regions
where the observed data are lower than 90\% of the sum of the
continuum and emission line models, and refit the spectra. The
deficit may result from absorption-lines.
\end{description}

\begin{figure}[tbp]
\epsscale{1.} \plotone{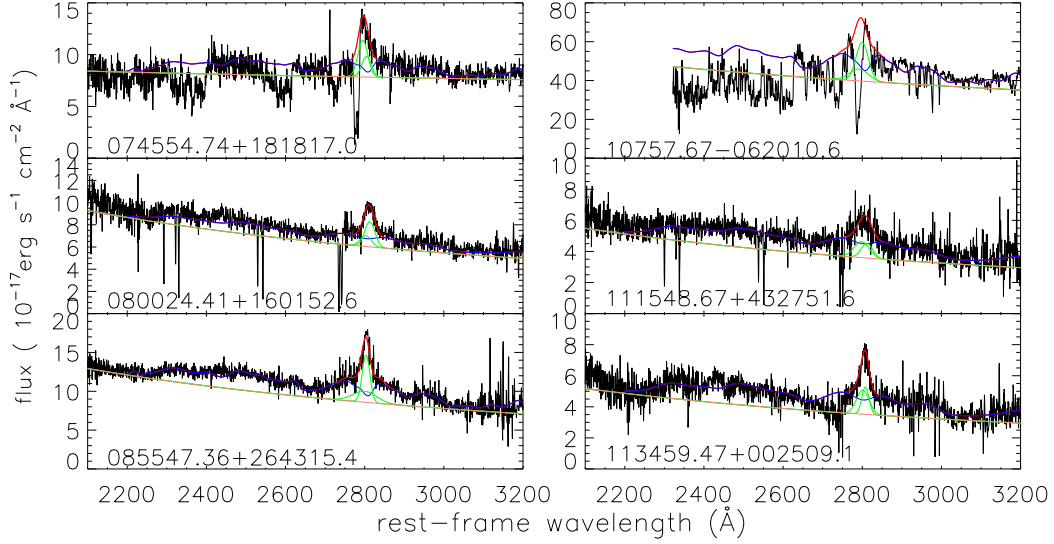} \caption{Observed spectra
overplayed with the best-fit models for the six quasars in the high-$z$
sample. The models are described in detail in \S2.1. In each panel,
we plot the observed spectrum in black curves, power-law continuum
in pink, broadened \ion{Fe}{2} template in blue, Gaussian \ion{Mg}{2}
emission line in green, and the model sum in red. } \label{f1}
\end{figure}

We do not include other UV emission lines, such as \ion{Fe}{3},
[\ion{Ne}{4}], and [\ion{C}{2}], which do not fall in the SDSS
wavelength coverage for many quasars in the low-$z$ sample. This
does not affect the measurement of \ion{Mg}{2} absorption line. The
fit is performed by minimizing $\chi^{2}$. The final model consists
of a single power law for the continuum, the empirical template
for UV \ion{Fe}{2} multiplets, and one or two Gaussian for
\ion{Mg}{2} emission line. It is worthy noting that our final model
does not include intrinsic dust extinction. We have tried to fit
spectra with reddening as a free parameter, and found that the
reddening and spectral continuum slope are degenerate. A similar
result has been found in Reichard et al. (2003a). As can be seen in
Fig. \ref{f1} (see also Fig. \ref{f3a}), this simple model can
reproduce the observed data very well for most quasars. We normalize
the observed spectra using the best-fit models, and measure
\ion{Mg}{2} absorption line, if present, in the normalized spectra.
We will validate in \S2.3 the reliability of our \ion{Mg}{2}
absorption line measurements by a careful comparison with previous
independent measurements that appeared in the literature.

\subsection{Criteria for \ion{Mg}{2} BAL Quasars}

Both of absorption-line width and depth vary from object to object,
and their distribution is broad and smooth. Upon that the fraction
of BAL quasars depends strongly on the definition. \ion{Mg}{2} is
more subtle than \ion{C}{4} because (1) loBALs are usually narrower than
HiBALs and (2) \ion{Mg}{2} is actually a doublet with a velocity
offset of 768 \kms. In this subsection, we will compare
different definitions that have appeared in the  literature and obtain a
more objective criterion of \ion{Mg}{2} BAL.

Weymann et al. (1991) proposed the first quantitative definition of BAL
by introducing the `balnicity index' (or $BI$ for short)
to describe BAL strength, which is defined as
\begin{equation}
BI = \int_{v_{l}=3,000}^{v_{u}=25,000}[1-{f(-v)\over 0.9}]C dv,
\label{eq:1.1}
\end{equation}
where $f$($-v$) is the normalized spectrum, and the velocity
$v$ in units of \kms~ with respect to the quasar emission-lines
(negative value indicates blueshift). The weight $C$ is set to 1  
when the observed spectrum falls at least 10\% below the model of
continuum plus emission lines in a contiguous velocity interval
($\Delta v_{c}$) of at least 2000 \kms, and $C$ = 0, otherwise. The
conservative value of 10\% is chosen to ensure that the deficit is
not caused by the uncertainty in the model of continuum and
emission lines. BAL quasars are defined as objects with $BI>0$ in at
least one absorption line.

\ion{Mg}{2} BAL troughs are generally weaker and narrower than those
found in \ion{C}{4}, and they usually show smaller blueshift than
\ion{C}{4} (Voit et al. 1993; Trump et al. 2006; G09).
Some bona fide BAL quasars would be lost if the same
`$BI$' definition is adopted for \ion{Mg}{2}. With this
consideration in mind, a few authors suggested to modify $BI$s by
changing the starting velocity and/or the interval of contiguous
absorption in the {\em Eq.}\ref{eq:1.1}. Hall et al. (2002)
redefined the starting velocity $v_{l}$ = 0 \kms~ and a minimum
contiguous interval of 450 \kms~ to give the `absorption
index' ($AI$ for short), and selected loBAL quasars according to the criterion
of $AI>0$ \kms. Trump et al. (2006) modified this definition
slightly so that $AI$ is a true equivalent width, measuring all
absorption within the limits of every absorption trough. Their $AI$
is defined as
\begin{equation}
AI  = \int_{v_{l}=0}^{v_{u}=29,000}[1-f(-v)]C' dv, \label{eq:1.2}
\end{equation}
where $C'$ = 1 when $\Delta v_c>$1000 \kms
at a depth of $>$10\% continuum; otherwise, $C'=0$.

Tolea et al. (2002) found that $BI$ distribution is very broad, and
there is no bimodality for BAL quasars in the SDSS early data
release. They pointed out that the definition is subjected to
certain arbitrariness (see also Weymann 2002). However,
Knigge et al. (2008) claimed that logarithmic $AI$ shows a bimodal
distribution  and argued that $BI$ works fairly well. Similarly, G09
introduced a modified $BI$, namely $BI_{0}$ by expanding the
integral from $v_{l}$=0 \kms, and they calculated both $BI$ and $BI_0$
for each ion (\ion{Si}{4}, \ion{C}{4}, \ion{Al}{3}, and \ion{Mg}{2}).
We set a maximum integration velocity of \ion{Mg}{2} BALs to 20,000 \kms~
to avoid confusion of \ion{Fe}{2} resonant absorption lines because
\ion{Mg}{2} BALs rarely extend to such a large velocity.
So, in this work the upper and lower limits on velocity for $AI$ integral
are 20,000 \kms~ and 0 \kms.
The definitions for BAL quasar are summarized in Table \ref{tab1}.

Since all known loBAL quasars show HiBALs, we can check consistency
of different definition of loBAL with that of HiBAL quasars based on
the presence of \ion{C}{4} BAL. With both  \ion{Mg}{2} and
\ion{C}{4} measurable, the high-$z$ quasar sample is well suited for
this purpose. We compare the fraction of \ion{C}{4} HiBAL, which is
defined as $BI_0>0$ (G09) among the loBAL quasar candidates with
various $\Delta v_{c}$ cutoff in \ion{Mg}{2} $AI$ definition.
We search for \ion{Mg}{2} BAL using different $\Delta v_c$, from 800
to 2000 \kms with an interval of $\delta v$= 200 \kms. The numbers
of \ion{Mg}{2} BAL quasar candidates are 395, 152, 119, 93, 73, 49 and 40,
for $\Delta v_{c}\geq$ 800, 1000, 1200, 1400, 1600, 1800 and 2000 \kms,
respectively. We cross-match these candidates with the HiBAL quasar
catalog of G09, and find that the fraction of HiBAL quasars in each
bin between two successive widths, $p(\Delta v_{i},
\Delta v_{i}+\delta v)$, increases with increasing $\Delta v_{i}$.
Specifically, all but three loBAL quasar candidates with $\Delta v_{c}\geq 1800$
\kms~ show \ion{C}{4} BALs, and about 79\% of the candidates do for
$1600\leq \Delta v_{c}<1800$ \kms. The percentage drops quickly as the
velocity cutoff decreases, for example $p(1200,1400)\sim 42\%$
and  $p(1000,1200)\sim 27\%$.
For $800\leq \Delta v_{c}<1000$ \kms~, only
$p\sim 15\%$ of the \ion{Mg}{2} BAL quasar candidates show
\ion{C}{4} BAL, which is similar to the fraction of HiBAL quasars
found in optically selected samples.

Assuming an average \ion{C}{4} BAL fraction of 15\% for optically-selected
quasars, we can estimate the true fraction of \ion{Mg}{2} BAL quasars in each bin
approximately as follows:

\begin{eqnarray}
\nonumber
x(\Delta v_{0}+i\delta v,\Delta v_{0}+(i+1)\delta v)=
\frac{p(\Delta v_{0}+i\delta v,\Delta v_{0}+(i+1)\delta v)-0.15}{1-0.15},\\
~~~~~~~~for~ i=0,1,2,3...
\end{eqnarray}
where $\Delta v_{0}=800$~\kms.

\begin{figure}[tbp]
\epsscale{0.6} \plotone{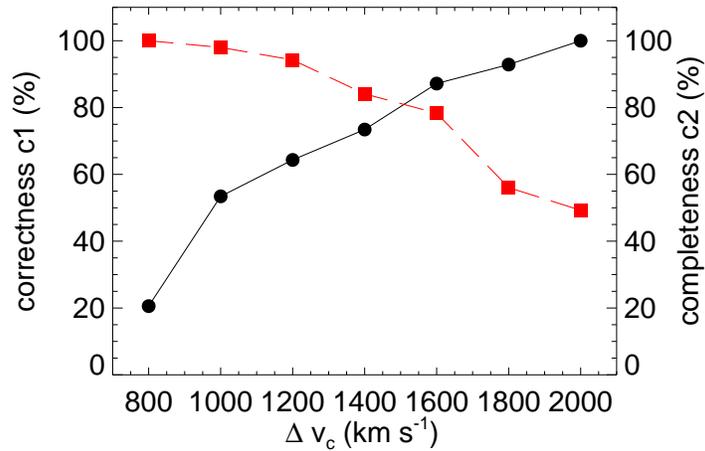} \caption{Correctness $c1$ (black
filled circles) and completeness $c2$ (red filled squares) of BAL
quasars as a function of \ion{Mg}{2} absorption-line width cutoff
$\Delta v_{c}$.} \label{f2}
\end{figure}

The `correctness' $c1$ and 'completeness' $c2$ as a function of
\ion{Mg}{2} absorption-line width cutoff $\Delta v_{c}$ can be estimated
as follows,

\begin{equation}
c1(\Delta v_{c}\geq \Delta v_{i})=\frac{\sum_{j=0}^{\infty}
x(\Delta v_{i}+j\delta v,\Delta v_{i}+(j+1)\delta v)\times
N(\Delta v_{i}+j\delta v,\Delta v_{i}+(j+1)\delta v))}
{N(\Delta v_{c}\geq \Delta v_{i})},
\end{equation}
\begin{equation}
c2(\Delta v_{c}\geq \Delta v_{i})=\frac{\sum_{j=0}^{\infty}
x(\Delta v_{i}+j\delta v,\Delta v_{i}+(j+1)\delta v)\times
N(\Delta v_{i}+j\delta v,\Delta v_{i}+(j+1)\delta v)}
{\sum_{j=0}^{\infty}x(\Delta v_{0}+j\delta v,\Delta v_{0}+(j+1)\delta v)
\times N(\Delta v_{0}+j\delta v,\Delta v_{0}+(j+1)\delta v)}
\end{equation}
where $N$ is the number of \ion{Mg}{2} BALs in the width bin between
two neighbor width cutoffs and a reasonable criterion should
be well balanced between `correctness' and `completeness'. The
cutoff $\Delta v_{i}\approx 1600$ \kms~ serves as a good trade-off
between the 'correctness' and 'completeness' (see Fig. \ref{f2}).
Using such a definition criterion, we have a correctness of
$c1(\Delta v_{c}\geq 1600~$\kms$)\approx 87.10\%$ and a completeness
of $c2(\Delta v_{c}\geq 1600~$\kms$)\approx 78.30\%$. We have
visually examined the \ion{Mg}{2} absorption-line profile, and found
that it very often splits into doublet for $\Delta v_{c}< 1600$
\kms, while it generally shows a smooth profile for $\Delta v_{c}
\geq 1600$ \kms. This also suggests that $ \Delta v_{c}\geq 1600$
\kms~ is a natural choice for defining loBAL quasars.

\subsection{Reliability of \ion{Mg}{2} Absorption-Line Measurements
and BAL Classifications}

Adopting above criterion, we culled 68 low-$z$ \ion{Mg}{2} BAL
quasars. The observed spectra and fits are displayed in Fig.
\ref{f3a}, and the properties of absorption and emission lines are
summarized in Table \ref{tab2}. In this subsection,  we examine the
reliability of our \ion{Mg}{2} absorption-line measurements and BAL
classifications by comparing with G09.

In the low-$z$ sample, 49 quasars are classified as \ion{Mg}{2} BAL
quasars by G09 according to the criterion of $BI_{0,Mg\;II}>0$.
Forty-one of them also fulfill our new criterion (marked with "B" in Fig. \ref{f3a}).
In Fig. \ref{f4}, we compare our measurements of
$EW_{absor}$, $V_{min}$ and $V_{max}$ with those of G09.
%For most quasars, they are very correspondency.
For most quasars, our measurements are the same as G09.
Large discrepancies in the minimum velocity are found for three quasars
(J1012+4921, J1044+3656 \& J1321+0202). Those quasars show two absorption
troughs, and G09 measured the larger velocity one, while we consider both
troughs (Fig 3). Significant difference in $V_{max}$ is found in
three quasars J1259+1213, J2107-0620 and J1321+0202.
In the former two, \ion{Mg}{2} BALs are potentially affected by \ion{Fe}{2}
absorption lines, and  our $AI$ integral is up to 20,000 \kms~ only,
so the measured $V_{max}$ are less than these of G09.
The third one is likely caused by different
continuum modeling: G09 did not include UV \ion{Fe}{2} model,
while we do.

We also find that eight \ion{Mg}{2} BAL quasars in G09's sample do not fulfill
our new criterion. Fig. \ref{f3a} shows the observed spectra and our
best-fit models for  these 'lost' \ion{Mg}{2} BAL quasars (marked
with "L"). We divide them into three cases: (1) no absorption line.
We do not detect absorption troughs at the level at least 10\%
below the best fitted continuum and emission line model in three objects.
\ion{Mg}{2} falls near the edge of SDSS spectral coverage for two
of them (J1015+3915 and 1018+5726), while the third object (J1142+0709)
show strong \ion{Fe}{2} emission lines. G09 did not include UV
\ion{Fe}{2} in their mod3el and \ion{Mg}{2} emission line is fitted with
Voigt profile; while we properly take into account  UV \ion{Fe}{2} and
adopt two Gaussian for \ion{Mg}{2} emission line. Different model should
account for the difference in the BAL classification in the last case.
(2) narrow absorption line. We measure only narrow absorption troughs
in other four objects (J0848+0345, J1051+5250, J1400+3539 and J1426+4112).
The difference is attributed to different approach
of continuum and emission line modeling. (3) J0802+5513. This object
displays a redshifted \ion{Mg}{2} BAL, relative to the
systematic redshift ($z = 0.663 \pm0.001$) measured from NELs,
such as [\ion{O}{2}] and [\ion{O}{3}], thus main part of
\ion{Mg}{2} BAL falls out the integration range of absorption
line index. With $V_{min}$=19,935 \kms~ and $V_{max}$=24,466 \kms,
G09 probably took the \ion{Fe}{2} absorption trough as the \ion{Mg}{2} BAL.

In our loBAL quasar sample, there are 27 objects which are not
included in G09's sample, we also show their spectra marked with "A"
in Fig. \ref{f3a}. More than half of them (15) have absorption line
width $ 1600\leq \Delta v_c< 2000$ \kms, while nearly a half sources
have $\Delta v_c\geq 2000$ \kms~ including five quasars (J0911+4035,
J1133+1112, J1250+4021, J1632+4204 and J1652+2158) $\Delta v_c >> 2000$
\kms. Thus the difference between our sample and G09 is caused
by the continuum and emission line modeling, as well as the
different definition of \ion{Mg}{2} BAL. Noting that most loBAL quasars
with $1600\leq \Delta v_c< 2000$ \kms~ show smooth, broad and deep absorption
troughs, just like loBAL quasars selected by $\Delta v_{c} \geq 2000$ \kms,
except for two sources, 1129+4228 and 1339+1119.
Therefore most of them should be \ion{Mg}{2} BAL quasars, and
our measurements should  be as reliable as G09. 

\subsection {Measurement of Emission Line and Continuum Parameters}
Our low-$z$ sample of \ion{Mg}{2} BAL quasars enable us to explore
the properties of many optical emission lines.
We measured in the SDSS spectra the parameters of broad and narrow
optical emission lines, including H$\gamma$, H$\beta$, \ion{Fe}{2},
[\ion{Ne}{5}]$\lambda$ 3425, [\ion{O}{2}]$\lambda$ 3728,
[\ion{Ne}{3}]$\lambda$ 3869, [\ion{O}{3}] $\lambda$$\lambda$ 4959,
5007 as well as the  continuum slope and normalization. The procedure of
the fitting is described in detail in Dong et al. (2008), and we will
only briefly outline it here. The optical continuum from 3600\AA~
to 5300 \AA~ is approximated by a single power-law. \ion{Fe}{2}
multiplets, both broad and narrow, are modeled using
the I Zw 1 templates provided by V{\'e}ron-Cetty et al. (2004).
Emission lines are modeled as multiple Gaussians: % one narrow Gaussians and
at most four Gaussians for broad  H$\beta$, one or two Gaussians for
[\ion{O}{3}]\footnote{The [\ion{O}{3}]$\lambda$5007 profile often
shows an extended blue wing and a sharp red falloff (Heckman et al
1981). We fit each of the [\ion{O}{3}] doublets with two Gaussians
when the blue wing is significant. We set upper-limits of 400 \kms~
and 500 \kms~ to the line shift and the $\sigma$ of the broad
Gaussian, respectively.}, and one Gaussian for each of the other 
NELs. We assume that the [\ion{O}{3}] doublet have the
same redshift and profile, and fix their doublet ratio
[\ion{O}{3}]4959/5007 to its theoretical value.
The equivalent widths of broad and narrow optical \ion{Fe}{2}~
multiplets in the composite spectra are calculated as
follows: $EW_{OptFeII}=\int_{4200\AA}^{5800\AA} ~f_{OptFeII}(\lambda) /
f_{con}(\lambda) d\lambda$, where, $f_{OptFeII}$ is the flux of broad
or narrow optical \ion{Fe}{2} emission, $f_{con}$ is the continuum flux.

%We use four parameters to describe H$\beta$ profile, namely FWHM,
%`sharpness', `asymmetry', and `boxiness'. The sharpness parameter,
%literally the departure from a triangle, is defined as $\frac{FW3/4M
%+ FW1/4M}{2FWHM}$, where FW1/4M, FWHM and FW3/4M are the line widths
%at 1/4,1/2,3/4 of the maximum intensity, respectively. The asymmetry
%parameter, defined as $(\lambda_{3/4}-\lambda_{1/4})/FWHM$, is a
%measure of the relative skewness of the peak and wings of the line.
%Both parameters were firstly introduced by Boroson \& Meyers (1992).
%The boxiness parameter is defined as ${FW3/4M}\over{FW1/4M}$, and is
%used to describe the `fatness' of line profiles (whittle 1985).

We measure the continuum slope $\beta_{[3K,4K]}$
($F_{\lambda} \propto \lambda^{\beta_{[3K,4K]}}$)
between $\sim 3000$ \AA~and $\sim 4000$ \AA~
for all quasars in the low-$z$ sample. The two continuum windows,
[3010, 3040]\AA~ and [4210, 4332]\AA, are so chosen to avoid strong
\ion{Fe}{2} multiplets shortward of 3000 \AA, and possible star-light
contribution longward of 5000 \AA~in some quasars. It is worthy to
mention that these bands are still affected by the Balmer continuum,
and significant \ion{Fe}{2} emission, though small. The slope thus may not
represent correctly the underlying intrinsic continuum in individual
quasar. However, it can be used as an indicator of continuum reddening
in a statistical way.

Using the measured continuum and emission line parameters, we
estimate black hole mass $M_{BH}$, using empirical relation between
black hole mass and the continuum luminosity and broad line width, and
Eddington ratio $L_{bol}/L_{Edd}$ for all quasars.
%, where $L_{Edd}\equiv
%\frac{4\pi Gcm_p}{\sigma_e}M_{BH}\approx 1.26\times
%10^{38}\frac{M_{BH}}{M_{\odot}}$ is the Eddington
%luminosity\footnote{$G$ is the gravitational constant, $c$ the
%velocity of light in vacuum, $m_p$ the mass of proton, and
%$\sigma_e$ the Thomson scattering section.},
The presence of
H$\beta$ emission line is important for the black hole mass estimate
because BAL associated with \ion{Mg}{2} often introduces a large
uncertainty in the measurement of \ion{Mg}{2} line width.
The black
hole mass is estimated using the following prescription (Vestergaard
\& Peterson 2006):
%which uses the optical continuum  luminosity and the H$\beta$ FWHM:
\begin{equation}
\log \,M_{\rm BH} (\rm H\beta) =
\log \,\left[ \left(\frac{\rm FWHM(H\beta)}{1000~km~s^{-1}}
\right)^2 ~
\left( \frac{\lambda \it L_{\lambda} {\rm (5100\,\AA)}}{10^{44}
\rm erg~s^{-1}}\right)^{0.50}
\right] + (6.91 \pm 0.02).
\label{Mbh_L51.eq}
\end{equation}
where, FWHM(H$\beta$) is the full width at half-maximum of
H$\beta$, after subtracting the narrow component. The bolometric
luminosity $L_{bol}$ is estimated from the monochromatic luminosity
at 5100\AA, $\lambda L_{\lambda}(5100$\AA$)$, with a bolometric
correction of 9 (Kaspi et al. 2000).

\section{RESULT AND ANALYSIS}
\subsection{Dust Extinction in \ion{Mg}{2} BAL Quasars}

It is known that that BAL quasars in general have redder continua
than non-BAL quasars, and loBAL quasars are even redder than HiBAL
quasars (Weymann et al. 1991; Brotherton et al. 2001; Reichard et
al. 2003b; Trump et al. 2006; G09). The red color of BAL quasars is
usually ascribed to the dust reddening in the BAL direction. We plot
the probability distributions of $\beta_{[3K,4K]}$ in
Fig.\ \ref{f5} panel (a) for both of
\ion{Mg}{2} BAL quasars and non-\ion{Mg}{2} BAL quasars.
Their distributions are both skewed to the red. The skewness
in the non-BAL quasars is likely due to dust reddening (Richards et
al. 2003). Although the width of the distributions for BAL and
non-BAL quasars are quite similar, BAL quasars are much redder than
non-BAL quasars. In fact, all but two \ion{Mg}{2} BAL quasars have
$\beta_{[3K,4K]}>-2.2$, which is the median value for non-BAL quasars.
The larger $\beta_{[3K,4K]}$ values of \ion{Mg}{2} BAL quasars are very
likely due to excess dust reddening in loBAL quasars as will be
discussed in the next section. Interestingly,% enough,
seven \ion{Mg}{2} BAL quasars have very red color of $\beta_{[3K,4K]} >
-0.2$, which form the red peak in Fig.\ \ref{f5} panel (a). We
visually inspected their observed spectra and found that this
results from significant contribution of starlight of their
host galaxies. Follow-up optical spectroscopy with a high $S/N$
ratio will be able to reveal the properties of their host galaxies.

We further compare the broad band SED 
between \ion{Mg}{2} BAL and non-\ion{Mg}{2} BAL quasars, from
ultraviolet through optical to near-infrared in quasar rest-frame.
We cross-correlate our low-$z$ quasar sample against the 2MASS point
source catalog (Skrutskie et al. 2006), and found 1993 matches
within $1^{''}$ offset between the optical and near-infrared
positions. Among them, 33 (or 1.7\%) objects are \ion{Mg}{2} BAL
quasars.
As can be seen in Fig.\ \ref{f6}, the rest-frame near-infrared to
optical colors of \ion{Mg}{2} BAL and non-BAL quasars are
essentially the same, while the former are significantly redder than
the latter in the ultraviolet.
This result is consistent with the interpretation
of the red color of loBAL quasars as dust reddening.

\subsection{Composite Spectra of \ion{Mg}{2} BAL Quasars}

To compare the average properties of non-\ion{Mg}{2}  BAL quasars
and \ion{Mg}{2}  BAL quasars, we created a geometric mean
(composite) spectrum for each class, following Vanden Berk et al.
(2001). For each quasar, we measure its redshift from 
NELs and deredshift the spectrum using the measured
redshift. The spectrum is then normalized at 3000 \AA, rebinned into
the same wavelength grids, and geometrically averaged bin by bin.
The composite spectra are created from the low-$z$ quasar sample, and
are shown in Fig.\ \ref{f7}(a).
\ion{Mg}{2} BAL quasars are redder than the non-BAL quasars. The
redness of \ion{Mg}{2} BAL composite spectrum is due to the overall
red SED instead of the BAL absorption troughs. To the first order
approximation, the composite spectrum of \ion{Mg}{2} BAL quasars is
similar to the non-\ion{Mg}{2} BAL quasars on both ultraviolet
\ion{Fe}{2} around \ion{Mg}{2}, H$\beta$, [\ion{O}{2}]$\lambda$3727,
[\ion{Ne}{3}]$\lambda$3869, and the red side of \ion{Mg}{2} line
profile, after de-reddening with an $E(B-V)=0.078$ for the SMC-like
extinction curve (blue curve in the panel (a) of Fig. \ref{f7}).
This extinction value is similar to that
obtained by Reichard et al. (2003b). However, there are subtle
differences between the two composite spectra. The \ion{Mg}{2} BAL
composite spectrum shows stronger
%\ion{Fe}{2} around 2070\AA,
optical \ion{Fe}{2}, but weaker [\ion{Ne}{5}]$\lambda$3425 in
comparison with non-\ion{Mg}{2} BAL composite spectrum. A deficit of
the \ion{Mg}{2} flux in the \ion{Mg}{2} BAL composite spectrum
starts well from the redside of the \ion{Mg}{2} line centroid. There
is also excess emission in the H$\beta$ wings in the \ion{Mg}{2} BAL
composite spectrum.
%The BAL composite is more wiggled between 1600 to 1800\AA, probably due to strong [\ion{N}{3}]$1750$ and \ion{Fe}{2} 191 emission and lack a broad hump beneath HeII$\lambda$1640.

In order to look into more detail the spectrum of \ion{Mg}{2} BAL
quasars with different properties, we divide the sample into
different bins according to their location on the continuum
spectral-index versus [\ion{O}{3}] equivalent width diagram (Fig.\
\ref{f5} panel (b)). A total six bins are adopted, each with a
combination of blue ($\beta_{[3K,4K]} \leq -2.2$), flat
($-2.2<\beta_{[3K,4K]} \leq -1.0$), red ($\beta_{[3K,4K]}>-1.0$) in
the spectral slope and [\ion{O}{3}]-strong ($EW_{[O\;III]}\geq
20$\AA) and [\ion{O}{3}]-weak ($EW_{[O\;III]}<20$\AA). Composite
spectra have been constructed separately for \ion{Mg}{2} BAL and
non-\ion{Mg}{2} BAL quasars in each bin with proper number of
sources. Because there are only two \ion{Mg}{2} BAL quasars with
$\beta_{[3K,4K]}\leq-2.2$ in our sample, we will not show the result
for \ion{Mg}{2} BAL quasars in the blue bin. Similarly, only
three \ion{Mg}{2} BAL quasars fall in the bin with red and
[\ion{O}{3}]-strong spectrum, thus we will not consider this bin. As
a result, only three composite BAL spectra are built. These
composite spectra are shown in Fig.\ \ref{f7}, while emission line
parameters are measured and listed in Table \ref{tab3}.

By comparing the composite spectra of \ion{Mg}{2} BAL quasars with
different spectral indices and [\ion{O}{3}] equivalent widths, we
find: (1) [\ion{O}{3}]-weak \ion{Mg}{2} BAL quasars show a deficit
in the \ion{Mg}{2} flux much larger and also extending to more
redward than [\ion{O}{3}]-strong \ion{Mg}{2} BAL quasars. (2) Other
NELs are also much weaker in the [\ion{O}{3}]-weak
\ion{Mg}{2} BAL quasars. In particular, [\ion{Ne}{5}]$\lambda$3424
is almost completely absent. However, when normalized to
[\ion{O}{3}], the narrow optical \ion{Fe}{2} emission is much
stronger. (3) The overall spectra of red and flat \ion{Mg}{2} BAL
quasars are very similar except for the continuum slope and some
narrow line strength, which can be ascribed to a combination of the
dust extinction plus a star-light contribution in the
long-wavelength.

The non-\ion{Mg}{2} BAL composite spectra have higher $S/N$ ratios
than their correspondent \ion{Mg}{2} BAL composite spectra due to a
large number of available spectra in each group. As in \ion{Mg}{2}
BAL quasars, the other NELs of [\ion{O}{3}]-weak
non-\ion{Mg}{2} BAL quasars are weak as well, suggesting an overall
weakness in narrow lines. Indeed, when normalized to [\ion{O}{3}],
the equivalent widths of narrow lines are similar for
[\ion{O}{3}]-weak and [\ion{O}{3}]-strong objects, except for narrow
\ion{Fe}{2} lines. The former also displays stronger broad optical
\ion{Fe}{2} emission as already noticed in many previous works
(e.g., Boroson \& Green 1992, hereafter BG92).
%The difference in the [\ion{O}{3}]-weak and [\ion{O}{3}]-strong sequence 
%can be understood in an [\ion{O}{3}]-strength related eigen-vector (BG92).
The equivalent widths of narrow lines increase from red to blue
composite spectra for both [\ion{O}{3}]-strong and [\ion{O}{3}]-weak
groups. Stellar absorption lines are visible in the composite of red
quasars. The presence of prominent high order Balmer absorption
lines suggests a recent starburst in the host galaxies. The
differences in the red to blue sequences can be explained by
increasing dust extinction from the red to blue sequences. As
extinction to the nucleus increases, the continuum and broad lines
dim more than narrow lines, resulting in an apparent increase in the
equivalent width of NELs. Because both extinction
and star-light contamination makes the spectrum redder, the
equivalent width of [\ion{O}{2}] increases more relative to
[\ion{O}{3}]. This is verified with the composite spectra (see also
Table \ref{tab3}).

The \ion{Mg}{2} BAL and non-\ion{Mg}{2} BAL quasars in the
[\ion{O}{3}]-weak group have similar equivalent widths of
[\ion{O}{3}] and of  broad H$\beta$.
In the flat spectrum group, \ion{Mg}{2} BAL quasars have stronger
Balmer narrow lines and narrow optical \ion{Fe}{2} component,
weaker [\ion{Ne}{3}], and much weaker [\ion{Ne}{5}] than
non-\ion{Mg}{2} BAL quasars, but have  a similar [\ion{O}{2}]
equivalent width. We find a similar behavior for the red spectrum
group as well, in comparison with non-\ion{Mg}{2} BAL quasars.
%In order to examine the difference between [\ion{Ne}{5}]-strong and
%[\ion{Ne}{5}]-weak quasars, we also select [\ion{Ne}{5}] detected
%and [\ion{Ne}{5}] non-detected quasars from high $S/N$ ($S/N \geq
%20$) spectra of non-\ion{Mg}{2} BAL quasars and construct composite
%spectra separately for the two groups. The emission line parameters
%are also listed in the Table \ref{tab3}. Non-[\ion{Ne}{5}] quasars
%have stronger narrow Balmer lines, much stronger narrow optical
%\ion{Fe}{2} emission, and also stronger broad optical \ion{Fe}{2}
%emission than [\ion{Ne}{5}] quasars.

\subsection{The Fraction of \ion{Mg}{2} BAL Quasars}

There are 68 \ion{Mg}{2} BAL quasars in the low-$z$ quasar sample,
which constitute 0.96\% of the sample. After correcting for the
missing and mis-classified BAL quasars introduced by $\Delta v_{c}$
cut in $AI$ discussed in \S 2.2, we find that the fraction of
\ion{Mg}{2} BAL quasars is $F_{Mg\;II\;BAL}=1.17\% $ in the SDSS
quasars. This fraction may  still under-estimate the true value
slightly because \ion{Mg}{2} BAL trough may fall outside of SDSS
spectral coverage for some quasars with $z \sim 0.4$. This number is
consistent with previous studies, which all yield a fraction of
about ~1\%  for \ion{Mg}{2} BAL quasars based on the visual
examination of \ion{C}{4} BAL quasars for the presence of strong
\ion{Mg}{2} absorption lines (Weymann et al. 1991; Boroson \& Meyers
1992; Turnshek et al. 1997). More recently, Trump et al. (2006)
found a fraction of 1.31\% for broad \ion{Mg}{2} BAL quasars in the
SDSS DR3 quasars in the redshift range $0.5<z<2.15$ that satisfied
an $AI$ definition of $\Delta v_{c}=1000$ \kms. With their
definition, we found a slightly high value of
2.05\%. 

As we have shown, \ion{Mg}{2} BAL quasars are usually redder than
non-\ion{Mg}{2} BAL quasars as a whole and this can be attributed to
dust reddening in \ion{Mg}{2} BAL quasars. Dust reddening will
introduce two additional selection effects for \ion{Mg}{2} BAL
quasars. First, dust extinction will dim \ion{Mg}{2} BAL quasars,
and hence they tend to be missed in a magnitude limit quasar sample.
Using the average extinction of 0.078 mag of SMC-like dust derived
by  the comparison of the composite spectra of \ion{Mg}{2} BAL quasar and
non-\ion{Mg}{2} BAL quasars, we find that the average extinction
correction to the $i$-band magnitude for loBAL quasars in the
observed frame is between 0.25 and 0.33 mag for redshifts from 0.4 to
0.8. Because the quasar luminosity is fairly steep with a slope of
-3.1 (Richards et al. 2003), this correction will introduce a factor
of around 2. However, the fraction of \ion{Mg}{2} BAL quasars missed
due to extinction is sensitive not only to the average excess
extinction, but also the  extinction distribution. Using the average
extinction will under-estimate the true fraction.
Second, because most SDSS quasars in the redshift range of $z\sim
0.4-0.8$ are selected via their colors and point-like morphologies,
the reddening would move some quasars outside of color locus for
quasar candidates, and the extinction to the nucleus also make the
galaxy contribution more prominent, as such they might be missed due
to their extend morphologies. Thus the corrected fraction of
\ion{Mg}{2} BAL quasars should be $>2\%$.

Recent surveys in other bands
have suggested that the number of optically-selected quasars will
miss half of the total quasar population (Martinez-Sansigre et al.
2005; Alonso-Herrero et al. 2006; Stern et al. 2007), but most of
the missed quasars may be type-2 rather than type-1, thus may not
significant affect the fraction of \ion{Mg}{2} BAL quasars.
Dust-reddened quasars missed in optical surveys have been found in
other non-optical surveys, such as hard X-ray surveys (Polletta et
al. 2007), infrared surveys (Cutri et al. 2001; Lacy et al. 2004;
Glikman et al. 2004, 2007) and radio surveys (White et al. 2003). It
has been suggested that the fraction of loBAL quasars are much
higher among those dust reddened quasars, e.g., 32\% (Urrutia et al.
2008). A pilot study of near-infrared bright quasar candidates from
UKDISS suggests that the reddening quasars missed by SDSS probably
accounts no more than 30\% (Maddox et al., 2008; cf Glikman et
al. 2007). From these multiband observation, the upper limit of the
\ion{Mg}{2} BAL fraction is estimated to be $<7\%$. To summarize,
the fraction of \ion{Mg}{2} BAL quasars is from 2\% to 7\%.

It was suggested that BAL quasar fraction increases with continuum
luminosity (Ganguly et al. 2007). However, G09 pointed out that $S/N$
and luminosity are degenerate, and it is unclear if BAL quasars are
truly more luminous or are simply identified at higher $S/N$. To
examine whether such a luminosity dependent trend is also seen in
loBAL quasars, and break the degeneracy of the $S/N$-luminosity
dependence, we split our sample into low and high luminosity
groups. The division line is so chosen that the two group has the
same size. The average luminosities for the low and high groups are
$\lambda L_{\lambda}(5100\AA)=2.8 \times
10^{44}$ and $\lambda L_{\lambda}(5100\AA)=8.8 \times 10^{44} $
erg~s$^{-1}$, respectively. We further divide the sample into
three equal size $S/N$ bins, and calculate the loBAL fraction in
each bin. The results are shown in the left panel of Fig.\ \ref{f8}.
For comparison, we also show the result of the whole sample.
As one can see, the fraction of \ion{Mg}{2} BAL quasars decreases with
decreasing median spectral $S/N$ ratios, from 1.7\% for $S/N \thicksim 25$
to 0.8\% for $S/N \thicksim 15$ for the whole sample. A similar trend is
also seen in the high luminosity group. However, the increase can be solely
due to luminosity effect because quasars in the highest $S/N$
ratio bin are a factor of two more luminous than its neigbor bin even
for the high luminosity group. The constancy of BAL fraction $7<S/N<15$
is an indication that S/N ratio does not actually matter. For a given $S/N$
ratio, the BAL fraction in high luminosity group is twice of
that in low luminosity group.

However, in each $S/N$ bin, there is still a variation of $S/N$.
In order to break the degeneracy, we try to use logistic
regression (e.g., Fox 1997) to assess how the likelihood of a
quasar being classified as BAL or non-BAL quasars depends on the
luminosity at 5100\AA~ and/or $S/N$.
Logistic regression solves for the natural logarithm of the odd ratio
in terms of the variables as follows:
\begin{equation}
 \ln {{Pr(G=1|L,S/N)}\over{Pr(G=2|L,S/N)}} = \beta_0 + \beta_1 \ln (S/N) +
         \beta_2 \ln (\lambda L_{\lambda}(5100\AA)/10^{44})
\label{logit_eq}
\end{equation}
where $G=1$ when a quasar is classified as BAL and $G=2$ as non-BAL.
The logit as the logarithm of the ratio of the probabilities, is
just the logarithm of the BAL to non-BAL ratio. The coefficients 
are calculated using the Newton-Raphson method, and errors are estimated 
by bootstrapping. We calculated the standard deviation of the
coefficients with 1000 random subsets of the true data. In the
Monte-Carlo simulation, we simulate the continuum subtracted
spectrum around \ion{Mg}{2} regime using the error array provided by
the SDSS pipeline, and remeasure the absorption line parameters as
for the real spectrum, i.e., reclassify as BAL or non-BAL quasars,
for each object. The continuum luminosity is generated using the
model parameters and their uncertainties. The uncertainty in the $S/N$
ratio is very small and is not considered. We obtain
$\beta_0=-4.42\pm 0.74$, $\beta_1=-0.78\pm0.48$ and
$\beta_3=1.37\pm0.32$.  In other words, the fraction of \ion{Mg}{2}
BAL decreases with $S/N$ at 1.6$\sigma$ significance and increases
with luminosity at 4.3$\sigma$ significance. Therefore, the analysis
supports that luminosity is a far more important factor in
determining the BAL fraction while it is far less affected by the $S/N$
ratio.  We show in the right panel of Fig. \ref{f8}
$F_{Mg\;II\;BAL}$ as a function of the luminosity. Our results
strongly suggest that \ion{Mg}{2} BAL quasars are averagely more
luminous than non-\ion{Mg}{2} BAL quasars, consistent with Ganguly
et al. (2007). G09 did not find such trend in their sample. The
difference may be due to a relatively larger luminosity range of our
sample than G09 or that \ion{Mg}{2} BAL quasars depend more
strongly on quasar luminosity than HiBAL quasars.

Fig.\ \ref{f9} shows $F_{Mg\;II\;BAL}$ as a function of
$EW_{[O\;III]}$, the equivalent width of [\ion{O}{3}] $\lambda$5007.
The fraction decreases from 1.65\% at $EW_{[O\;III]}=8.6$ \AA~ to
0.51\% at $EW_{[O\;III]}=34.3$\AA. Because both $F_{Mg\;II\;BAL}$
and $EW_{[O\;III]}$ are correlated with optical luminosity (Dietrich
et al. 2002), there is a concern that  such a dependence is caused
by luminosity effect. In order to break the degeneracy, we split the
sample into two luminosity bins. Both the high and low luminosity
bins show the same trend with a higher overall fraction and  steep
slope  for high luminosity bin, thus dependence of \ion{Mg}{2} BAL
fraction on [\ion{O}{3}]-strength is not a secondary  effect of
luminosity.

Boroson (2002) argued that BAL quasars occupy the high Eddington
ratio and the black hole mass locus on the Eddington ratio versus
black hole mass diagram. Ganguly et al. (2007) found that the frequency
and properties of BALs depend on both  black hole mass and
Eddington ratio for their sample of HiBAL quasars. We also
investigate the possible dependence of $F_{Mg\;II\;BAL}$ on $M_{BH}$
and $L/L_{Edd}$. We divided our quasars into three bins in either
Eddington ratio or black hole mass, and calculate $F_{Mg\;II\;BAL}$
in each bin. We find that $F_{Mg\;II\;BAL}$ in the highest Eddington
ratio bin ($L/L_{Edd} \geq 0.24$) is significantly higher (two times)
than that in rest of the two bins, the difference is more than 2.6$\sigma$
%while there is no clear difference between the rest two bins
(Fig. \ref{f10}). However, we do not find any significant
correlation between $F_{Mg\;II\;BAL}$ and $M_{BH}$, in the range
from a few $10^7$ to $10^9$~$M_\odot$ in our sample. In passing, we
note that using other formalisms of black hole mass estimate based
on broad H$\beta$ line (e.g., Kaspi et al. 2000; Green \& Ho
2005; Wang et al. 2009) yields the same conclusion.

\subsection{Emission Line Properties of \ion{Mg}{2} BAL Quasars}

Based on a small sample of infrared selected quasars, Boroson \&
Meyers (1992) proposed that \ion{Mg}{2} BAL or loBAL quasars have
very weak even undetectable [\ion{O}{3}] emission compared to
non-loBAL quasars. In their sample, the [\ion{O}{3}] equivalent
widths of loBAL quasars are typically less than 5 \AA. As discussed
in the last section, we find that there are still a fraction of
loBAL quasars with a relatively high [\ion{O}{3}] equivalent width,
though the fraction of \ion{Mg}{2} BAL quasars in the
[\ion{O}{3}]-strong sample is only one-third of that in the
[\ion{O}{3}]-weak sample.
Only 20 of 68 \ion{Mg}{2} BAL quasars have their $EW_{[O\;III]}$ smaller
than 5 \AA, while 13 \ion{Mg}{2} BAL quasars have higher than 20 \AA, some of
them have even higher than 100 \AA. We compare the properties of
[\ion{O}{3}]-strong and [\ion{O}{3}]-weak \ion{Mg}{2} BAL quasars,
and find significant difference between the two types (Table \ref{tab3}).
First, [\ion{O}{3}]-strong loBAL quasars show much weaker \ion{Mg}{2}
absorption feature (Fig. \ref{f7} (b)). The median value of $AI$ is
631 \kms in the [\ion{O}{3}]-strong loBAL sub-sample, while the
median value of $AI$ is 1141 \kms~ in [\ion{O}{3}]-weak loBAL
sub-sample. Second, [\ion{O}{3}]-weak loBAL quasars also display
systematically weaker broad H$\beta$ line as shown in Fig \ref{f7}
(b). Third, [\ion{O}{3}]-weak loBAL quasars have strong UV and
optical \ion{Fe}{2} emission, and a stronger narrow component of
\ion{Fe}{2}, in particular. The median $EW_{UV\;Fe\;II}$ are
136.01$^{+12.79}_{-3.99}$ \AA~ and 113.6$^{+2.64}_{-1.38}$ \AA~
for [\ion{O}{3}]-weak and [\ion{O}{3}]-strong loBAL quasars, respectively.
%The parameters
%of optical \ion{Fe}{2} emission are listed in the Table \ref{tab3}.
The intensity ratio of broad optical \ion{Fe}{2} component to broad
H$_{\beta}$ component in the [\ion{O}{3}]-weak loBAL composite
spectrum is twice as strong as that in the [\ion{O}{3}]-strong
BAL composite spectrum, and the intensity ratio of
\ion{Fe}{2} narrow component to [\ion{O}{3}]
is more than 20 times larger in the former.

In order to find the connection between the outflow and emission
line region, we explore the correlation between the outflow
parameters and emission line parameters. We find that $AI$ is
marginally correlated with UV \ion{Fe}{2} (with a Spearman rank
correlation coefficient of $r_s=0.27$ and a null hypothesis
probability of $P_r=0.05$) in the sense that there is an upper
envelope of $AI$, which increases with $EW_{Fe\;II\;UV}$
(Fig.\ref{f11} (a)).  As shown in Fig.\ref{f11} (b), there seems
also an upper border in the $AI$ for a given $EW_{[O\;III]}$, which
decreases with $EW_{[O\;III]}$ and there seems also a low boundary
for $AI$. $AI$ is marginally anti-correlated $EW_{[O\;III]}$ with
$r_s=-0.24$ and $P_r=0.07$. Apparently, larger sample are required to
confirm these weak trends.

\subsection{Properties of Intermediate Width \ion{Mg}{2} Absorption
            Line Quasars}

In \S 2.2, we defined our criteria of \ion{Mg}{2} BAL quasars by
comparing the frequency of \ion{C}{4} BAL with $AI(\Delta v_{c}\geq2000)>0$ at
different cuts in $\Delta v_{c}$ of \ion{Mg}{2} absorption lines. We finally
choose a contiguous absorption with a minimum depth of 10\% of the
intrinsic emission spectrum over a width $\Delta v_{c}=1600$ \kms. With
this definition, we can identify BAL quasars with \ion{Mg}{2}
absorption lines that are associated with BALs. Note that Trump et
al. (2006) and some other previous works have used 1000 \kms~ as the
selection criteria for loBAL quasars. If quasars with absorption
width between 1000 \kms~ and 1600 \kms~ are true \ion{Mg}{2} BAL,
they should have similar properties in the emission lines and continuum
as these selected with $\Delta v_c= 1600$ \kms. In the following, we will
check whether there is a clear distinction in the \ion{Mg}{2} absorbers
with a width above and below 1600 \kms.

In our low-$z$ quasar sample, there are 68 \ion{Mg}{2} BAL quasars with
$\Delta v_{c}\geq1600$ \kms, and  76 quasars with $1000\leq
\Delta v_{c}<1600$\kms~(hereafter lowV group). We compare continuum and
emission line properties of these two groups with non-\ion{Mg}{2}
BAL quasars as well as between the two groups either statistically
or using their composite spectra. Table \ref{tab3} summarizes the
emission line parameters measured from the composite spectrum of
lowV quasars. We find that the emission line properties, including
\ion{Fe}{2}, [\ion{Ne}{5}] and [\ion{O}{3}] of lowV quasars are
similar to non-\ion{Mg}{2} BAL quasars, but  different
from BAL quasars. However, the composite spectrum of lowV group is
significantly redder than that of non-\ion{Mg}{2} BAL quasars.

We also measure the parameters of broad and narrow optical emission
lines and continuum, including H$_{\beta}$ profile, the equivalent
width of UV \ion{Fe}{2} and [\ion{Ne}{5}], FWHM of [\ion{O}{3}] and
[\ion{O}{2}], $\lambda L_{\lambda}(5100$\AA$)$ and $\beta_{[3K,4K]}$
for each quasar. Fig. \ref{f12} shows the distributions of line and
continuum parameters for the two samples of \ion{Mg}{2} absorbers
as well as that of non-\ion{Mg}{2} BAL quasars.
Significant difference between lowV quasars and \ion{Mg}{2} BAL
quasars is found in distributions of following parameters: the
equivalent width of UV \ion{Fe}{2}, [\ion{Ne}{5}] and $\lambda
L_{\lambda}(5100$\AA$)$, as shown in Fig.\ref{f12}.
Kolmogorov-Smirnov (KS) test gives null probabilities of $9\times10^{-3}$,
$6\times10^{-3}$ and $6\times10^{-5}$, respectively, for
UV \ion{Fe}{2}, [\ion{Ne}{5}] and $\lambda L_{\lambda}(5100$\AA$)$,
that their distributions are drawn from the same parent population for lowV
and \ion{Mg}{2} BAL quasars.  \ion{Mg}{2} BAL quasars have
weaker [\ion{Ne}{5}] and [\ion{O}{2}], stronger optical \ion{Fe}{2} and
higher observed optical luminosity
than lowV  quasars. The two groups have almost the same distributions
in continuum slope.
The lowV quasars show the similar distributions of all emission line
parameters as non-\ion{Mg}{2} BAL quasars, except for a redder continuum
and lower observed optical luminosities.

\section{Summary and Discussion}

We have selected 68 \ion{Mg}{2} BAL quasars from the SDSS
spectroscopic quasar sample with a redshift of $0.4<z\leq0.8$ and a
median $S/N \geq 7$ using criteria of a continuous absorption over a
velocity interval greater than 1600 \kms~for a depth of at least
10\%. The BAL-selected criterion is a trade-off between the
completeness and consistency with respect to the canonical
definition of BAL quasars that have the 'balnicity index' $BI>0$ in
\ion{C}{4} BAL. We adopted such a criterion to ensure that $\sim
90\%$ of our sample are classical BAL quasars and the completeness
is $\sim 80\%$, based on extensive tests using high-$z$ quasar
samples with measurements of both \ion{C}{4} and \ion{Mg}{2} BALs.
The low-$z$ sample is used to define the fraction of \ion{Mg}{2} BAL
quasars and its dependence on the continuum and emission line
properties, the difference between \ion{Mg}{2} and non-\ion{Mg}{2}
BAL quasars. We find that, (1) the fraction of \ion{Mg}{2} BAL
quasars in the optical survey is around 1.2\%. The fraction does not
include correction for internal dust extinction of BAL quasars and
the color bias against reddened loBAL quasars. After correcting
these factors, the true \ion{Mg}{2} BAL fraction is likely in
between 2\% and 7\%. (2) \ion{Mg}{2} BAL quasars are more frequently
found in quasars with low [\ion{O}{3}] equivalent width and high
continuum luminosity although they show a wide range of [\ion{O}{3}]
equivalent width. loBAL quasars display stronger narrow optical
\ion{Fe}{2} emission lines and UV \ion{Fe}{2} emission, weaker even
absent [\ion{Ne}{5}] lines. (3) The fraction of quasars with
\ion{Mg}{2} BAL increases strongly with the Eddington ratio but does
not correlate with the black hole mass. (4) There is an excess of
intrinsic reddening in \ion{Mg}{2} BAL quasars and quasars with
intermediate width \ion{Mg}{2} absorption lines with an average of
0.08 mag for SMC-like dust grain. In this
section, we will discuss the implication of our results.%its implication.

It is generally believed that NELs are nearly
isotropic in quasars because they are produced in an extended region.
In contrast the BEL region and continuum are thought
to be much more compact and can be blocked on some lines of sight
by obscuration (e.g., Antonucci 1993).
The excess extinction in \ion{Mg}{2} BAL quasars with respect to
non-\ion{Mg}{2} BAL quasars will enhance the conclusion that
[\ion{O}{3}] equivalent width is lower in loBAL quasars.
The conclusion will be further strengthened if we consider the
anisotropic emission of optical continuum from the accretion disk
because it is generally believed that BAL quasars are seen nearly
edge-on. The large range of observed $EW_{[O\;III]}$ among \ion{Mg}{2}
BAL quasars suggests that loBAL occurs in both [\ion{O}{3}]-weak and
[\ion{O}{3}]-strong emission quasars, but the covering factor
decreases strongly as the [\ion{O}{3}] strength increases.

If [\ion{O}{3}] is considered as an indicator of overall strength of
NELs, the strength of other lines relative to
[\ion{O}{3}] should connect more to the physical state of 
narrow-line region (NLR) or
ionizing continuum. The absence of [\ion{Ne}{5}] in
[\ion{O}{3}]-weak \ion{Mg}{2} BAL turns out to be a rather surprise.
Previous studies have suggested that [\ion{Ne}{5}] is produced in
the high density, inner NLR (e.g., Heckman et al. 1981; De Robertis
\& Osterbrock 1984; Whittle 1985a, 1985b). Lack of [\ion{Ne}{5}]
emission indicates that there is no such region or the inner NLR
does not expose to a hard ionizing continuum. Interaction of BAL
outflow with inner NLR may destroy dense clouds in the inner NLR.
However, the presence of strong narrow optical \ion{Fe}{2} emission
would suggest such dense inner NLR does exist but with low
ionization parameters (see V{\'e}ron-Cetty et al. 2004; also Wang,
Dai \& Zhou 2008). Then we look at the option that the NLR only sees
a soft ionizing continuum. BAL quasars, loBAL quasars in particular,
are weak in soft X-rays (Green et al. 1995; Brinkmann et al. 1999),
which are required to produce  Ne$^{4+}$ (ionization potential 97
eV). If NLR sees a continuum similar to the observed one, the
absence of [\ion{Ne}{5}] can be naturally explained. Because the
weakness of soft X-rays in BAL quasars is usually attributed to
X-ray absorption rather than intrinsic weakness (Wang et al. 1999;
Gallagher et al. 1999, 2002), this requires that the [\ion{Ne}{5}]
emission region is behind the X-ray absorber. There are two
possibilities for this, the outflow has a large covering factor
or [\ion{Ne}{5}] emission region located coincidently  behind the
outflow. Nagao et al. (2001) proposed that [\ion{Ne}{5}] emission
region is the inner region of the dust torus, and use the hypothesis
to explain the unification of two type Seyfert galaxies. In their
scenarios, [\ion{Ne}{5}] emission region lies on the equatorial
plane, which is coincident with the region shielded by disk wind.

In order to check this, we select 250 non-\ion{Mg}{2} BAL quasars
with $S/N >20$, $\beta_{[3K,4K]} > -2.2$ and $EW_{[O\;III]} < 20$,
out of which 119 quasars do not show detectable [\ion{Ne}{5}]
emission line in the spectra. Fig.\ \ref {f13} shows the two
composite spectra of these non-\ion{Mg}{2} BAL quasars with/without
[\ion{Ne}{5}] emission, the emission parameters are shown in Table
\ref{tab3}. The composite spectrum of the 119 non-BAL quasars
without [\ion{Ne}{5}] is similar to that of \ion{Mg}{2} BAL quasars,
showing stronger optical narrow \ion{Fe}{2} emission, weak
[\ion{O}{3}] strength and strong UV \ion{Fe}{2} but with a blue
continuum. These quasars are probably from the same parent
population of loBAL quasars but our line of sight does not intersect
the outflow. The  high fraction of non-BAL quasars without
[\ion{Ne}{5}] emission line indicates
that the covering factor of loBALR is not large.
%relative high frequency of [\ion{Ne}{5}] indicates
%that the covering factor of BALR is not large.

Several observed trends may be explained by the strong correlation
between the frequency of loBAL and Eddington ratio, and the
correlations of the Eddington ratio with the other parameters concerned.
It was reported that narrow optical \ion{Fe}{2} strength is fairly
well correlated with the Eddington ratio for low redshift quasars (Dong
et al. 2009b). Dietrich et al. (2002) demonstrated that [\ion{O}{3}]
strength is inversely correlated with the Eddington ratio for quasars.
Weakness of \ion{Mg}{2} in the red-side of \ion{Mg}{2} line profile
is difficult to be ascribed to the absorption, and can be understood
in this context as well via a fairly strong anti-correlation between
EW of \ion{Mg}{2} and the Eddington ratio (Dong et al. 2009a). Thus,
the Eddington ratio can be an underlying driver for the different
covering factor of low ionization BALR. We note that even in the
highest Eddington ratio bin, the fraction of loBAL quasars is only a
factor of 2   of the value in the rest two low Eddington ratio
bins. There is no correlation between the fraction of loBAL with
black hole mass for this sub-sample. Narrow line Seyfert 1 galaxies
(NLS1s) are believed to be the low mass counter-parts of high
accretion rate quasars. Zhou et al. (2006a; 2006b) found that several
NLS1s also show \ion{Mg}{2} BALs.
Twelve of our low-$z$ \ion{Mg}{2} BAL quasars can be formally classified as
NLS1s according to the formal criterion of $H\beta <2000~km~s^{-1}$,
which account for 1.6\% of NLS1s in this redshift range. Therefore,
NLS1s do not appear to show significantly different properties from
other quasars with  similar optical luminosity. The black hole mass
range of these loBAL-NLS1s is very narrow with [2.1, 7.8] $\times 10^7$
M$_{\sun}$. It is possible that black hole mass does not matter once
it is above certain threshold. %We will discuss in detail the possible
%connection between loBAL and NLS1 phenomena in a separated paper.

Finally, most quasars with \ion{Mg}{2} BALs or intermediate width
\ion{Mg}{2} absorption lines show reddened colors. We already
noticed that the two group quasars show significantly different
properties of emission lines, and the BAL fraction among the latter
group is less than 25\% (refer \S 2.2).
Thus, it is unlikely due to mixing of un-identified BAL quasars.
The ubiquity of dust in intermediate and LoBAL outflows may be naturally
explained as both absorbers are large scale outflows (e.g., Dunn et al.
2010). The presence of dust will significantly boost the radiative force
and thus it allow gas in a relative large distance from the nucleus to be
accelerated by the quasar radiation. In other word, gas free from dust will
not be accelerated to high velocity by radiation pressure. In  this case,
dust reddening would be preferably observed in the outflow direction.
As we have argued that [\ion{Ne}{5}]-weak quasars may be \ion{Mg}{2}
BAL quasars seen from an off-BALR direction, and their color can be fairly blue,
thus dust may not present in other direction. This is consistent with above argument. 
Certainly, critical test for this can be done with a comparative study of
broad band infrared SED of BAL and non-BAL quasars.

Dust reddening is ubiquitous in broad ($\Delta v_{c} \geq 1000$ \kms)
\ion{Mg}{2} absorbers, regardless of whether they are BAL or non-BAL quasars.
The average excess reddening is E(B-V) $\sim$ 0.08 mag for SMC-type
dust for both groups. However, we think that the intermediate
width \ion{Mg}{2} absorption quasars might have somewhat lower
reddening than \ion{Mg}{2} BAL quasars. Because quasars with large
extinction, more likely BAL quasars, are missed in the SDSS quasar
sample due to color selection criteria of quasar target, this sample
explores only the relative low extinction end of quasars.  Thus the
similar color distribution for quasars with \ion{Mg}{2} BAL and with
intermediate width \ion{Mg}{2} absorption lines may be caused by the
color selection effect that introduces a truncation in the severely
reddened quasars. Indeed,
radio and infrared-selected \ion{Mg}{2} BAL quasars show much larger
extinction with E(B-V) up to 1.5 mag (Urrutia et al. 2008), while
there is no good statistical work for intermediate width
\ion{Mg}{2}. So it is not conclusive whether \ion{Mg}{2} BAL quasars
and quasars with intermediate width \ion{Mg}{2} absorption lines
have similar dust extinctions.

\acknowledgements

This work has made use of the data obtained by SDSS. Funding for the
SDSS and SDSS-II has been provided by the Alfred P. Sloan
Foundation, the Participatings Institutions, the National Science
Foundation, the U.S. Department of Energy, the National Aeronautics
and Space Administration, the Japanese Monbukagakusho, the Max
Planck Society, and the Higher Education Funding Council for
England. The SDSS Web Site is http://www.sdss.org/.

The SDSS is managed by the Astrophysical Research Consortium for the
Participating Institutions. The Participating Institutions are the
American Museum of Natural History, Astrophysical Institute Potsdam,
University of Basel, University of Cambridge, Case Western Reserve
University, University of Chicago, Drexel University, Fermilab, the
Institute for Advanced Study, the Japan Participation Group, Johns
Hopkins University, the Joint Institute for Nuclear Astrophysics,
the Kavli Institute for Particle Astrophysics and Cosmology, the
Korean Scientist Group, the Chinese Academy of Sciences (LAMOST),
Los Alamos National Laboratory, the Max-Planck-Institute for
Astronomy (MPIA), the Max-Planck-Institute for Astrophysics (MPA),
New Mexico State University, Ohio State University, University of
Pittsburgh, University of Portsmouth, Princeton University, the
United States Naval Observatory, and the University of Washington.

\clearpage

\begin{deluxetable}{lcccl}
 \tabletypesize{\tiny}
 \tablecaption{Different Definition for the BAL Quasars
 \label{tab1} }
 \tablewidth{0pt}
 \tablehead{
 \colhead{} &
 \colhead{$v_{l}$} &
 \colhead{$v_{u}$} &
 \colhead{$\Delta v_{c}$} &
 \colhead{Reference}
}
 \startdata
 $BI$ &3,000  &25,000 &2,000  &Weymann et al. 1991\\
    &       &       &       &Tolea et al. 2002\\
    &       &       &       &Reichard et al. 2003a\\
    &       &       &       &Trump et al. 2006\\
    &       &       &       &G09\\
 $BI$ &0      &25,000 &1,000  &Reichard et al. 2003a\\
 $BI_0$ &0  &25,000 &2,000  &G09\\
 $AI$ &0    &25,000 &450    &Hall et al. 2002\\
 $AI$ &0      &29,000 &1,000  &Trump et al. 2006\\
 $AI$ &0      &20,000 &1,600  &This work\\
 \enddata
\end{deluxetable}

\begin{deluxetable}{lrrrrrrrrrrrrrrr}
 \tabletypesize{\tiny}
 \rotate
 \tablecaption{Low-$z$ \ion{Mg}{2} BAL Quasars Catalog in SDSS DR5
 \label{tab2} }
 \tablewidth{0pt}
 \tablehead{
 \colhead{Name$^{a}$} &
 \colhead{R.A.} &
 \colhead{Decl.} &
 \colhead{Redshift} &
 \colhead{AI} &
 \colhead{$deepth_{max}^{b}$} &
 \colhead{$V_{max}^{c}$}&
 \colhead{$V_{min}^{d}$}&
 \colhead{$V_{ave}^{e}$}&
 \colhead{$\lambda L_{\lambda}(5100$\AA$)$}&
 \colhead{$\beta_{[3K,4K]}$}&
 \colhead{$EW_{[O\;III]}$}&
 \colhead{$EW_{[Ne\;V]}$}&
 \colhead{$M_{BH}$}&
 \colhead{$L/L_{EDD}$}\\
 \colhead{} &
 \colhead{J2000} &
 \colhead{J2000} &
 \colhead{} &
 \colhead{\kms} &
 \colhead{} &
 \colhead{\kms} &
 \colhead{\kms} &
 \colhead{\kms} &
 \colhead{$10^{44}~$erg s$^{-1}$} &
 \colhead{} &
 \colhead{\AA} &
 \colhead{\AA}&
 \colhead{$10^8~M_{\odot}$}&
 \colhead{}
 }
\startdata
  010352.46+003739.7 &   15.968604 &    0.627704 & 0.7031 &    480$\pm$   1 & 0.73 & 11500 &  9100 & 10280 &  22. & -1.21 &  15.1 &   0.6 &  16. &  0.097\\
  023102.49-083141.2 &   37.760410 &   -8.528133 & 0.5868 &    303$\pm$   2 & 0.72 &  3750 &  2100 &  2966 & 3.5 & -2.04 &  18.5 &   0.9 &  2.1 &  0.12\\
  025026.66+000903.4 &   42.611093 &    0.150945 & 0.5963 &   1032$\pm$   8 & 0.32 &  3400 &   750 &  2066 &  5.4 &  0.99 &  24.2 &   3.6 &  37. &  0.010\\
  080934.64+254837.9 &  122.394340 &   25.810552 & 0.5454 &    519$\pm$   1 & 0.49 &  1600 &     0 &   665 &  9.2 & -2.11 &  13.4 &   0.4 &  18. &  0.036\\
  081655.34+074311.5 &  124.230604 &    7.719886 & 0.6442 &    312$\pm$   2 & 0.75 &  5500 &  3750 &  4703 &  5.7 & -1.52 &  18.1 &   3.9 &  2.6 &  0.15\\
  082231.53+231152.0 &  125.631381 &   23.197803 & 0.6530 &    677$\pm$   1 & 0.42 &  2350 &   450 &  1346 &  18. & -1.62 &  51.0 &   2.2 &  50. &  0.027\\
  083525.98+435211.2 &  128.858253 &   43.869796 & 0.5678 &    781$\pm$   2 & 0.66 & 20000 & 16250 & 18120 &  13. & -1.58 &   4.3 &   0.0 &  6.0 &  0.16\\
  083613.23+280512.1$^{\dag}$ &  129.055150 &   28.086698 & 0.7412 &  4105$\pm$  11 & 0.30 & 17050 &  5800 & 10546 &  5.3 & -1.51 &   7.0 &   0.0 &  0.73 &  0.52\\
  084716.04+373218.0 &  131.816838 &   37.538359 & 0.4539 &    347$\pm$   2 & 0.73 &  4100 &  2400 &  3265 &  3.5 & -2.22 & 156.6 &   4.5 &  0.98 &  0.26\\
  085215.66+492040.8 &  133.065270 &   49.344685 & 0.5664 &   1143$\pm$   4 & 0.14 &  2250 &     0 &  1202 &  4.3 & -1.50 &  10.7 &   0.7 &  4.1 &  0.076\\
  085357.87+463350.6 &  133.491147 &   46.564063 & 0.5497 &    418$\pm$  1 & 0.62 &  4400 &  2800 &  3642 &  8.2 & -1.47 &   3.5 &   0.2 &  6.2 &  0.095\\
  091146.06+403501.1$^{\dag}$ &  137.941942 &   40.583652 & 0.4412 &   1053$\pm$  7 & 0.33 &  6100 &    3150 &  4082 &  2.9 & -0.09 &  21.2 &   2.4 &  0.41 &  0.51\\
  092157.62+103539.0$^{\dag}$ &  140.490091 &   10.594188 & 0.5476 &   1730$\pm$   8 & 0.49 &  8950 &  3250 &  6068 &  2.7 & -1.61 &   7.4 &   0.5 &  0.33 &  0.58\\
  092928.63+324129.9 &  142.369302 &   32.691639 & 0.7748 &    279 $\pm$  2 & 0.79 &  9900 &  8050 &  8997 &  1.3 & -1.78 &   9.3 &   0.2 &  5.3 &  0.18\\
  093034.79+570520.7 &  142.644964 &   57.089097 & 0.6374 &    314  $\pm$ 2 & 0.72 & 10350 &  8750 &  9493 &  6.8 & -2.00 &   7.1 &   0.0 &  1.7 &  0.29\\
  094443.13+062507.4 &  146.179709 &    6.418734 & 0.6951 &   1532$\pm$   1 & 0.48 &  9300 &  4300 &  7245 &  69. & -1.42 &   2.9  &   0.0 &  18. &  0.28\\
  101203.31+492148.2 &  153.013803 &   49.363391 & 0.7389 &   1030$\pm$   3 & 0.56 &  7650 &   700 &  4029 &  8.5 & -1.73 &   1.1 &   0.0 &  15. &  0.041\\
  102021.21+121909.1 &  155.088391 &   12.319199 & 0.4793 &   3320$\pm$  17 & 0.31 & 18350 &     0 & 11309 &  4.0 & -0.05 &   2.4 &   0.0 &  1.0 &  0.28\\
  102802.32+592906.6 &  157.009676 &   59.485190 & 0.5349 &    995$\pm$   3 & 0.36 &  2650 &   300 &  1421 &  2.5 & -2.11 &   7.1 &   0.9 &  2.5 &  0.072\\
  102839.11+450009.4 &  157.162961 &   45.002620 & 0.5826 &    773$\pm$   1 & 0.41 &  2450 &   100 &  1204 &  24. & -1.41 &  11.8 &   0.0  &  8.9 &  0.20\\
  103621.60+393701.6 &  159.090011 &   39.617129 & 0.7997 &    777$\pm$   1 & 0.48 &  3100 &   200 &  1488 &  40. & -1.31 &   5.0 &   0.0 &  48. &  0.059\\
  104210.43+501609.2 &  160.543497 &   50.269224 & 0.7873 &   1384$\pm$   1 & 0.13 &  3650 &  1050 &  2503 &  20. & -2.23 &   1.1 &   1.8 &  21. &  0.070\\
  104459.60+365605.1 &  161.248363 &   36.934775 & 0.7015 &   3578$\pm$   2 & 0.02 & 17650 &   800 &  6311 &  42. & -1.34 &   1.7 &   0.0 &  9.1 &  0.33\\
  105259.99+065358.1$^{\dag}$ &  163.249983 &    6.899474 & 0.7224 &   2508$\pm$   7 & 0.53 & 12900 &  4600 &  8792 &  8.9 & -1.27 &   9.8 &   0.0 &  0.69 &  0.92\\
  112632.81+430938.6 &  171.636726 &   43.160726 & 0.4356 &    886$\pm$   3 & 0.24 &  1800 &     0 &   874 &  6.2 & -0.05 &   5.2 &   0.7 &  3.9 &  0.11\\
  112730.71+423039.0 &  171.877963 &   42.510858 & 0.5310 &    391$\pm$   4 & 0.52 & 19050 & 17300 & 18079 &  3.3 & -1.81 &  67.3 &   4.9 &  0.75 &  0.31\\
  112822.41+482309.9 &  172.093398 &   48.386109 & 0.5428 &    370$\pm$   1 & 0.57 &  4450 &  2850 &  3565 &  16. & -0.83 &   2.7 &   0.0 &  15. &  0.077\\
  112912.27+422853.9$^{\dag}$ &  172.301146 &   42.481651 & 0.5805 &    308$\pm$   3 & 0.74 &  6250 &  4500 &  5362 &  4.6 & -1.16 &  15.4 &   0.0 &  0.57 &  0.58\\
  113355.22+111208.8 &  173.480118 &   11.202465 & 0.7612 &    706$\pm$   4 & 0.65 & 10250 &  5400 &  7829 &  6.4 & -1.91 &   4.5 &   0.0 &  0.98 &  0.46\\
  113807.83+531231.7 &  174.532647 &   53.208815 & 0.7899 &    843$\pm$   3 & 0.36 &  4700 &  2750 &  3790 &  1.1 & -1.38 &   5.8 &   0.0 &  3.5 &  0.23\\
  114043.62+532439.0 &  175.181765 &   53.410835 & 0.5304 &    886$\pm$   4 & 0.37 &  3650 &  1350 &  2612 &  4.5 & -1.69 &  17.9 &   3.3 &  8.4 &  0.039\\
  114915.30+393325.4$^{\dag}$ &  177.313775 &   39.557076 & 0.6284 &   1291$\pm$   7 & 0.46 &  5050 &  1600 &  3326 &  4.2 & -1.25 &  23.9 &   1.2 &  0.33 &  0.90\\
  115816.72+132624.1$^{\dag}$ &  179.569678 &   13.440048 & 0.4389 &    949$\pm$   2 & 0.60 &  8650 &  4750 &  6484 &  7.7 & -1.62 &  18.2 &   0.1 &  0.65 &  0.85\\
  121113.38+121937.3 &  182.805763 &   12.327049 & 0.4641 &   1441$\pm$   4 & 0.18 &  3950 &   600 &  2076 &  3.1 & -0.78 &  24.4 &   0.0 &  2.1 &  0.11\\
  121303.40-014450.9 &  183.264183 &   -1.747485 & 0.6123 &    684$\pm$   3 & 0.37 &  4200 &  2350 &  3250 &  7.8 & -1.27 &  11.6 &   0.0 &  1.9 &  0.29\\
  122043.21-013215.3 &  185.180046 &   -1.537600 & 0.4478 &    915$\pm$   6 & 0.33 &  2100 &   100 &   990 &  1.8 & -1.41 & 160.5 &   6.5 &  12. &  0.011\\
  124300.87+153510.6 &  190.753663 &   15.586296 & 0.5609 &   1173$\pm$   2 & 0.16 &  1700 &     0 &   856 &  5.5 & -1.56 &  37.5 &   0.7 &  9.2 &  0.043\\
  125057.57+402100.3 &  192.739909 &   40.350086 & 0.6070 &   1789$\pm$   8 & 0.05 &  8350 &  2550 &  4427 &  3.9 & -1.25 &   3.0 &   1.2 &  1.2 &  0.24\\
  125942.80+121312.6$^{\dag}$ &  194.928337 &   12.220168 & 0.7487 &   2716$\pm$  10 & 0.49 & 13400 &  2250 &  7892 &  12. & -0.57 &  14.7 &   0.0 &  0.93 &  0.94\\
  130741.12+503106.5 &  196.921364 &   50.518476 & 0.6987 &   1714$\pm$   3 & 0.09 &  6250 &  2500 &  4127 &  13. & -0.81 &   0.61 &   0.0&  22. &  0.042\\
  131823.73+123812.6 &  199.598896 &   12.636836 & 0.5848 &   1804$\pm$   5 & 0.07 &  4150 &   700 &  2384 &  3.2 & -1.51 &   4.9 &   4.9 &  7.1 &  0.033\\
  132114.42+020225.0 &  200.310112 &    2.040294 & 0.5813 &    445$\pm$   4 & 0.58 &  5050 &  3200 &  4059 &  3.2 & -0.56 &  17.6 &   0.0 &  0.95 &  0.24\\
  133936.69+111949.2 &  204.902879 &   11.330361 & 0.6489 &    382$\pm$   4 & 0.65 &  7050 &  5300 &  6199 &  4.9 & -1.34 &   9.1 &   0.6 &  0.91&  0.39\\
  134415.75+331719.1$^{\dag\ddag}$ &  206.065639 &   33.288650 & 0.6856 &    771$\pm$   2 & 0.30 &  1800 &     0 &   901 &  7.7 & -1.28 & 112.4 &   5.2 &  0.82&  0.68\\
  135418.26+585935.9 &  208.576089 &   58.993326 & 0.7907 &    827$\pm$   5 & 0.38 &  3850 &  1750 &  2893 &  6.3 & -0.81 &  33.1 &   1.7 &  1.1 &  0.40\\
  140025.53-012957.0 &  210.106416 &   -1.499180 & 0.5839 &   1013$\pm$   5 & 0.38 & 20000 & 17400 & 18844 &  6.5 & -0.98 &   4.0 &   0.0 &  3.3 &  0.14\\
  142649.24+032517.7 &  216.705181 &    3.421588 & 0.5295 &    480$\pm$   3 & 0.63 &  3450 &  1600 &  2637 &  5.3 & -1.25 &   5.9 &   0.0 &  4.4 &  0.086\\
  142927.28+523849.5 &  217.363680 &   52.647085 & 0.5939 &   1009$\pm$   2 & 0.63 &  5900 &  2050 &  3957 &  15. & -1.25 &   8.2 &   0.0 &  2.9 &  0.38\\
  143303.27+403105.1 &  218.263643 &   40.518104 & 0.4452 &    680$\pm$   4 & 0.38 &  3350 &  1200 &  2324 &  2.4 & -1.80 &  15.9 &   1.2 &  1.9 &  0.089\\
  143828.63+452108.6 &  219.619318 &   45.352395 & 0.4268 &    324$\pm$   3 & 0.74 &  8250 &  6500 &  7367 &  3.2 & -1.95 &  18.8 &   0.3 &  0.67 &  0.34\\
  144642.26+454631.0 &  221.676097 &   45.775287 & 0.7364 &    675$\pm$   4 & 0.58 &  5450 &  2950 &  4198 &  12. & -1.09 &   0.0 &   0.9 &  4.2 &  0.20\\
  145724.01+452157.7$^{\dag}$ &  224.350045 &   45.366055 & 0.7175 &   1691$\pm$   5 & 0.62 & 14200 &  6800 & 10259 &  8.6 & -1.45 &   2.8 &   0.0 &  0.95 &  0.65\\
  145736.70+523454.6 &  224.402944 &   52.581834 & 0.6376 &   1194$\pm$   4 & 0.68 &  9150 &  3600 &  6041 &  11. & -1.74 &   8.7 &   0.2 &  2.4 &  0.34\\
  145836.73+433015.5 &  224.653068 &   43.504323 & 0.7595 &    615$\pm$   5 & 0.62 &  6500 &  3850 &  5122 &  9.7 & -1.58 &   4.7 &   0.5 &  6.5 &  0.11\\
  150847.41+340437.7 &  227.197577 &   34.077147 & 0.7880 &    722$\pm$   1 & 0.52 &  2850 &   200 &  1573 &  34. & -1.48 & 134.1 &   5.8 &  30. &  0.083\\
  152350.42+391405.2 &  230.960111 &   39.234787 & 0.6612 &   3227$\pm$   2 & 0.43 & 20000 & 11500 & 16458 &  38. & -1.14 &   4.8 &   0.1 &  39. &  0.069\\
  153036.82+370439.1 &  232.653452 &   37.077543 & 0.4170 &   1911$\pm$   5 & 0.44 &  8100 &  2000 &  4657 &  7.0& -1.20 &  13.5 &   0.3 &  1.2 &  0.44\\
  160329.72+502722.2 &  240.873867 &   50.456176 & 0.6389 &    762$\pm$   5 & 0.49 &  5300 &  2800 &  4133 &  4.5& -1.61 &  23.3 &   2.3 &  2.4 &  0.13\\
  160628.06+290333.8 &  241.616939 &   29.059398 & 0.4347 &    489$\pm$   3 & 0.63 &  3650 &  1500 &  2696 &  9.1 & -0.14 &   8.3 &   4.3 &  1.3 &  0.49\\
  160721.27+515510.6 &  241.838665 &   51.919619 & 0.7739 &    404$\pm$   3 & 0.63&  6800 &  5200 &  5909 &  6.9 & -2.19 &   4.7 &   2.3 &  0.88 &  0.57\\
  163255.46+420407.7 &  248.231106 &   42.068816 & 0.7263 &   1152$\pm$   4 & 0.56 &  8650 &   750 &  3985 &  7.8 & -1.53 &  18.2 &   4.7 &  39 &  0.014\\
  163513.51+213859.6$^{\dag}$ &  248.806315 &   21.649900 & 0.6835 &    573$\pm$   5 & 0.69 &  6950 &  4300 &  5620 &  4.9 & -1.57 &   6.3 &   0.3 &  0.61 &  0.58\\
  165225.39+215830.7 &  253.105807 &   21.975220 & 0.4468 &   1155 $\pm$ 12 & 0.42 &  7850 &  4450 &  6170 &  3.6 & -0.26 &  18.8 &   0.0 &  0.68 &  0.38\\
  170010.83+395545.8$^{\dag}$ &  255.045127 &   39.929392 & 0.5766 &    462 $\pm$  6 & 0.67 &  5550 &  3750 &  4594 &  3.7 & -0.69 &  12.5 &   1.2 &  0.51 &  0.52\\
  170341.82+383944.7 &  255.924255 &   38.662439 & 0.5539 &    785 $\pm$  7 & 0.43 &  5000 &  2750 &  3949 &  5.9 &  0.01 &   8.1 &   1.9 &  5.0 &  0.084\\
  204333.20-001104.2 &  310.888342 &   -0.184524 & 0.5449 &   2074 $\pm$  4 & 0.16 &  8250 &  2750 &  5061 &  8.9 & -1.12 &   7.7 &   0.0 &  2.0 &  0.31\\
  210757.67-062010.6 &  316.990293 &   -6.336294 & 0.6456 &   2344 $\pm$  1 & 0.21 &  8250 &     0 &  3709 &  27. &  0.01&  18.5 &   1.9 &  76. &  0.026\\
  224028.14-003813.1 &  340.117275 &   -0.636989 & 0.6591 &   1364 $\pm$  2 & 0.23 &  3650 &   750 &  1994 &  9.9 & -1.19 &   8.2 &   0.0 &  8.1 &  0.087\\
\enddata
\tablenotetext{a}{SDSS DR5 designation, hhmmss.ss+ddmmss.s
(J2000.0)} \tablenotetext{b}{Deepest part of any BAL trough}
\tablenotetext{c}{Maximum velocity of the BAL troughs from the
emission line} \tablenotetext{d}{Minimum velocity of the BAL troughs
from the emission line} \tablenotetext{e}{Weighted average velocity
of the BAL troughs}
\tablenotetext{\dag}{narrow-line Seyfert 1}
\tablenotetext{\ddag}{double-peaked narrow lines}
\end{deluxetable}

\begin{deluxetable}{lcccccccccccccc}
 \tabletypesize{\tiny}
 \rotate
 \tablecaption{Parameters of Emission Lines from Composite Spectra
  \label{tab3}}
 \tablewidth{0pt}
 \tablehead{
 \colhead{} &
 \colhead{N-B-H$^{a}$} &
 \colhead{N-B-L} &
 \colhead{M-F-H} &
 \colhead{N-F-H} &
 \colhead{M-F-L} &
 \colhead{N-F-L} &
 \colhead{N-R-H} &
 \colhead{M-R-L} &
 \colhead{N-R-L} &
 \colhead{no[Ne V]$^{b}$} &
 \colhead{[Ne V]$^{c}$} &
 \colhead{nonBAL} &
 \colhead{LowV$^{d}$} &
 \colhead{BAL$^{e}$}
}
 \startdata
$EW_{[O\;III]}$ (\AA) &38.0&12.4&58.0&36.8&8.7&10.7&38.4&5.8&10.7&8.1&19.5&18.4&21.9&14.2
\\
\hline
$[$O III]      &1     &1     &1     &1     &1     &1 &1     &1     &1     &1     &1    &1      &1      &1\\
H$\beta^N$ &0.051&0.035&0.060&0.068&0.098&0.16&0.071&0.11&0.098&0.26&0.19 & 0.084&0.058 &0.19\\
H$\gamma^N$ &0.030&0.035&0.041&0.039&0.083&0.079&0.032&0.046&0.036&0.11&0.12& 0.048&0.0792 &0.097\\
OptFeII$^N$ &0.16&0.24&0.088&0.057&3.4&1.8&0.13&1.7&0.67&4.8&0.52&0.41&0.45 &2.0\\
$[$Ne V] &0.065&0.091&0.070&0.077&0.017&0.089&0.081&0.0092&0.14&0.00070&0.089&0.083 &0.084 &0.051\\
$[$O II] &0.070&0.051&0.083&0.097&0.14&0.11&0.15&0.29&0.18&0.083 &0.078&0.14&0.11 &0.15\\
$[$Ne III] &0.086&0.12&0.067&0.099&0.098&0.16&0.12&0.29&0.21&0.18 &0.11&0.15&0.15 &0.10\\
\hline
$EW_{H\beta^B}$ (\AA) &117.7&75.7&89.9&74.9&68.4&61.1&60.3&57.5&50.6&66.5&78.4& 76.9&71.6 &69.5\\
\hline
H$\beta^B$  &1  &1  &1  &1  &1  &1 &1  &1  &1  &1  &1  &1  &1  &1\\
H$\gamma^B$ &0.28 &0.23 &0.24 &0.24 &0.25 &0.24 &0.26 &0.29 &0.21 &0.26 &0.21 & 0.25&0.23 &0.24\\
OptFeII$^{B}$ &0.77 &2.2 &0.79 &1.4 &3.5 &3.6 &1.4 &3.3 &3.2&3.7&2.4&2.2&2.2 &2.7
\enddata
 \tablenotetext{a}{Names of the composite spectra: X-Y-Z \\
 ~~~~~~X: N-- non-Mg II BAL  quasar;
 %~~~~~~X: N-- non-\ion{Mg}{2} BAL  quasar;
 M-- Mg II BAL quasar\\
 %B-- \ion{Mg}{2} BAL quasar\\
 ~~~~~~Y: B-- $\beta_{[3K,4K]} \leq -2.2$; F-- $-2.2 < \beta_{[3K,4K]} \leq -1$;
 R-- $\beta_{[3K,4K]} >  -1$\\~~~~~~Z: H-- $EW_{[O\;III]} \geq$ 20 \AA; L-- $EW_{[O\;III]} <$ 20 \AA}
 \tablenotetext{b}{The composite spectrum of non-Mg II BAL quasars with $S/N >$ 20 and undetectable [Ne V] emission}
 \tablenotetext{c}{The composite spectrum of non-Mg II BAL quasars with $S/N >$ 20 and detectable [Ne V] emission}
 \tablenotetext{d}{The composite spectrum of 1000 \kms $\leq$ $\Delta v_{c} <$  1600 \kms selected  quasars in low-$z$ quasar sub-sample}
 \tablenotetext{e}{The composite spectrum of $ \Delta v_{c}  \geqslant$  1600 \kms selected  Mg II BAL quasars in low-$z$ quasar sub-sample}
\end{deluxetable}
\clearpage

\figurenum{1}

\figurenum{2}

\figurenum{3}
\begin{figure}[tbp]
\epsscale{.9} \plotone{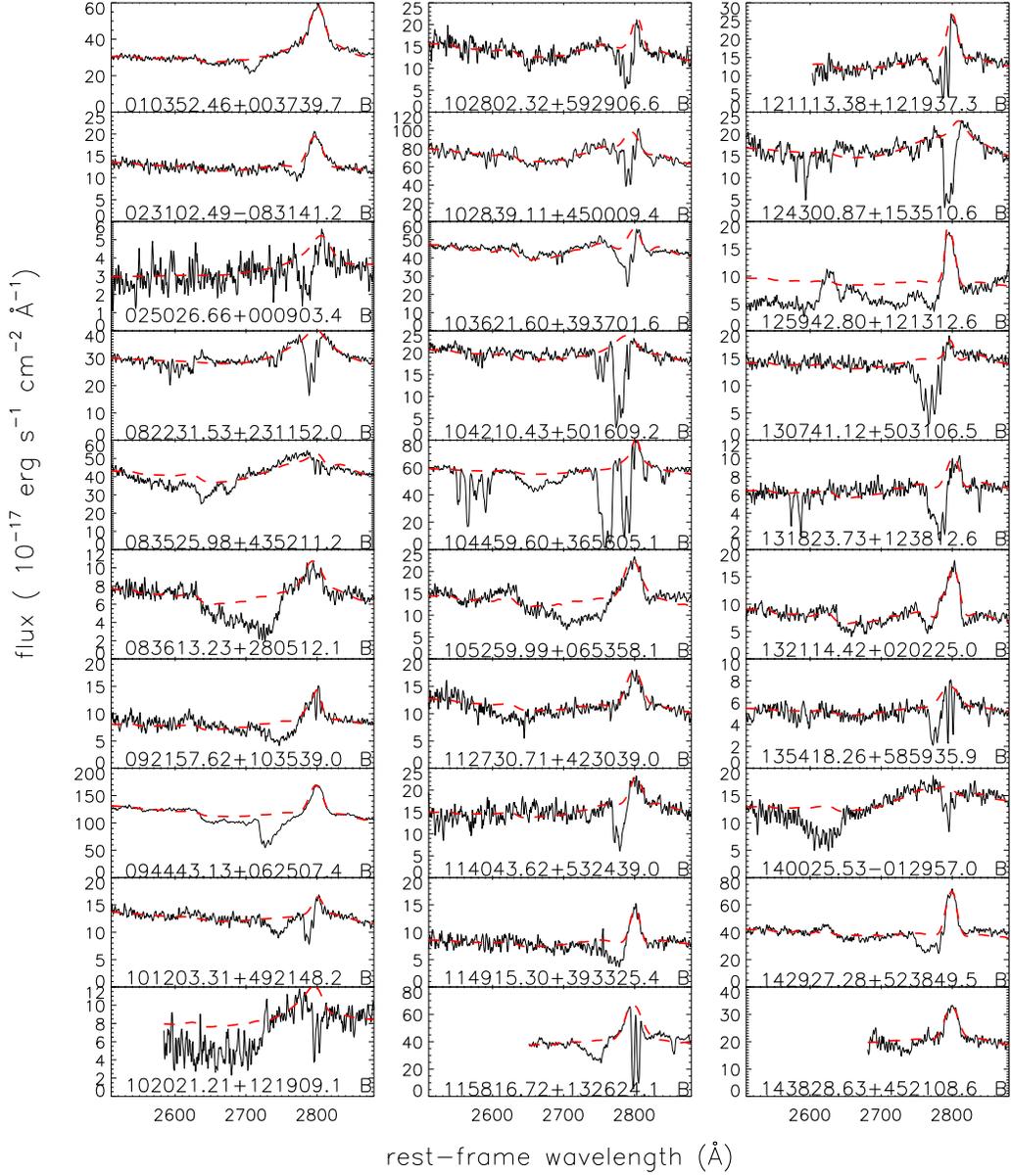} \caption{Observed spectra and
our best-fit models of the low-$z$ sample of 68 \ion{Mg}{2} BAL quasars,
including 41 classified as such by both of us and G09 (marked
with "B"), and 27 in this work but not in G09 (marked with
"A"). Also shown here are eight objects that are not classified
as \ion{Mg}{2} BAL quasars according to our new criterion, but are
classified as such in G09 (marked with "L"). }  \label{f3a}
\end{figure}

\figurenum{3}
\begin{figure}[tbp]
\epsscale{.9} \plotone{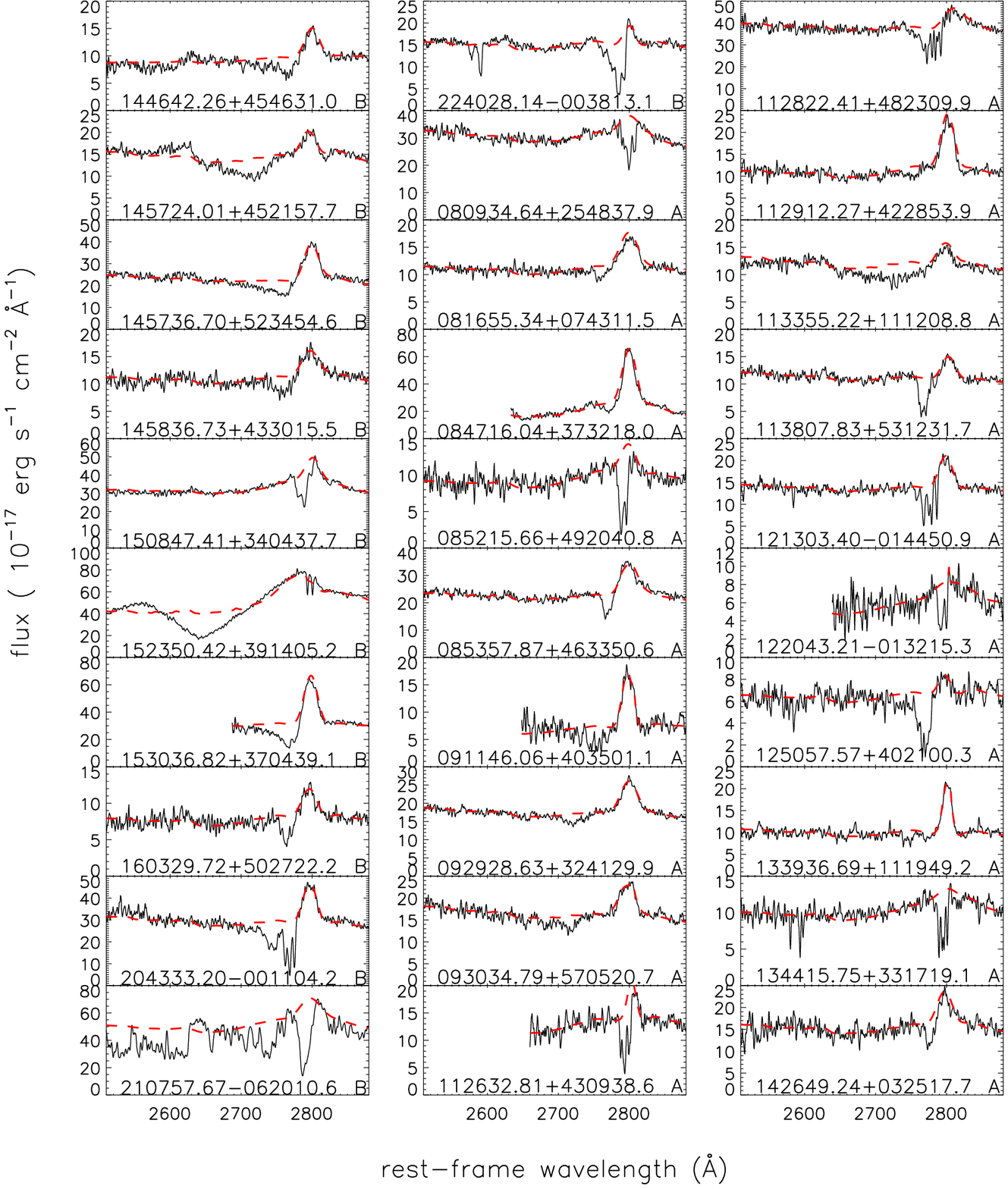} \caption{Continued... } \label{f3b}
\end{figure}

\figurenum{3}
\begin{figure}[tbp]
\epsscale{1.0} \plotone{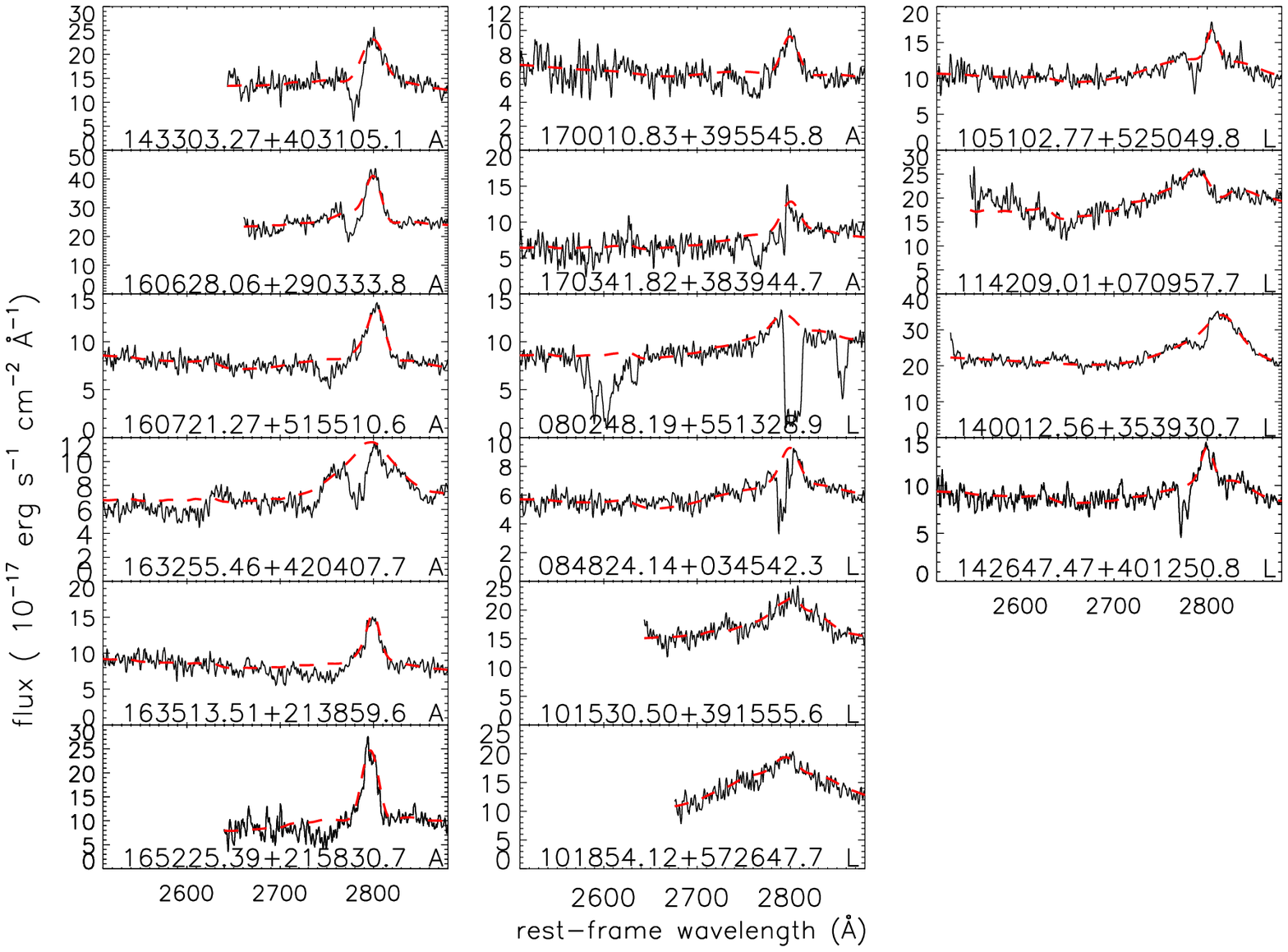} \caption{Continued. } \label{f3c}
\end{figure}
\clearpage

\figurenum{4}
\begin{figure}[tbp]
\epsscale{1.0} \plotone{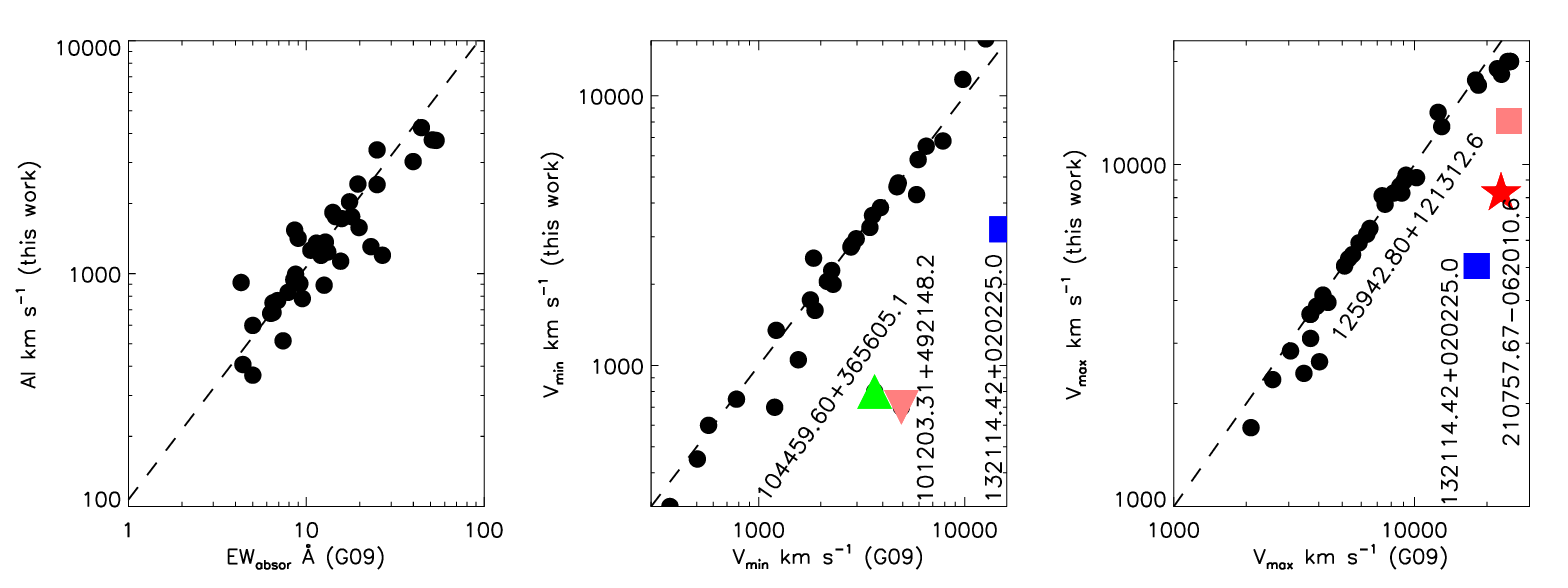} \caption{Comparison of \ion{Mg}{2}
BAL parameters between our measurements and G09's. The 41 objects in
the low-$z$ sample are classified as \ion{Mg}{2} BAL quasars in
this work and G09. \textit{Left:} absorption index `AI' vs.
rest-frame equivalent width, \textit{middle:} minimum velocity
$V_{min}$, and \textit{right:} maximum velocity $V_{max}$. The
dashed lines show one-to-one correspondence. Overall, our
measurements agree well with that of G09. Significant
differences are found for five quasars. The origin of the discrepancies
examined in \S2.3 and the observed spectra of these five quasars are
shown in Fig. 3, together with our best-fit models. } \label{f4}
\end{figure}

\figurenum{5}
\begin{figure}[tbp]
\epsscale{1.0} \plotone{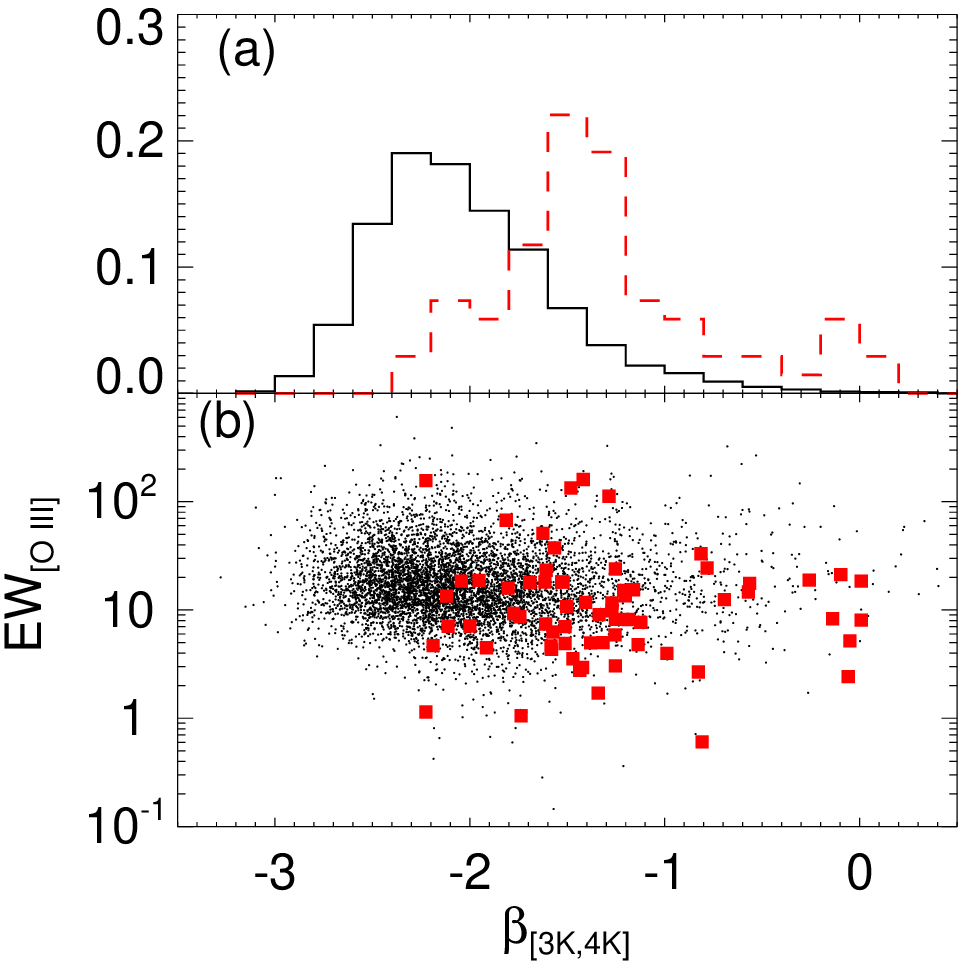} \caption{Panel (a): Comparison of the
distribution of continuum slope $\beta_{[3K,4K]}$ between
\ion{Mg}{2} BAL quasars (red dash line) and non-loBAL quasars
(black thick line). It is obvious that \ion{Mg}{2} BAL quasars are much
redder than non-loBAL quasars. The $\beta_{[3K,4K]}$ distribution of
non-\ion{Mg}{2} BAL quasars peaks at -2.2, much less than the value
of -1.7 for \ion{Mg}{2} BAL quasars. The median value of
$\beta_{[3K,4K]}$ is -2.08 for non-\ion{Mg}{2} BAL quasars, and it
is -1.41 for \ion{Mg}{2} BAL quasars. Panel (b): Equivalent width
of \ion{O}{3} NEL $EW_{[O\;III]}$ plotted against
continuum slope $\beta_{[3K,4K]}$ for the low-$z$ sample. \ion{Mg}{2}
loBAL quasars are shown in filled red squares, and non-loBAL quasars
in filled black circles.} \label{f5}
\end{figure}

\figurenum{6}
\begin{figure}[tbp]
\epsscale{1.0} \plotone{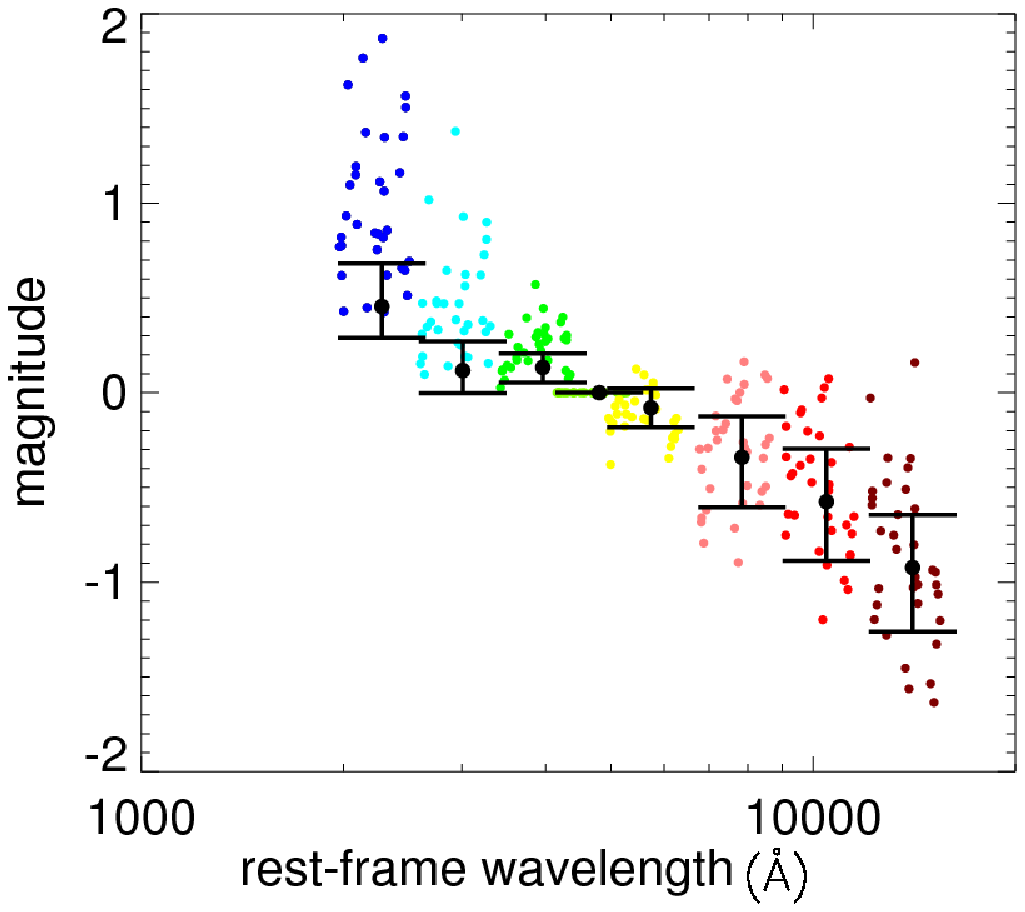}  \caption{Comparison of the
UV-optical-NIR SED of \ion{Mg}{2} BAL and
non-loBAL quasars. All the data are normalized at the SDSS $i$-band.
The different bands of \ion{Mg}{2} BAL quasars are shown by filled circle
in different colors. The black filled circles show the mean SED of
non-loBAL quasars, with the upper and lower quartile denoted by error
bars.} \label{f6}
\end{figure}
\clearpage

\figurenum{7}
\begin{figure}[tbp]
\epsscale{.8} \plotone{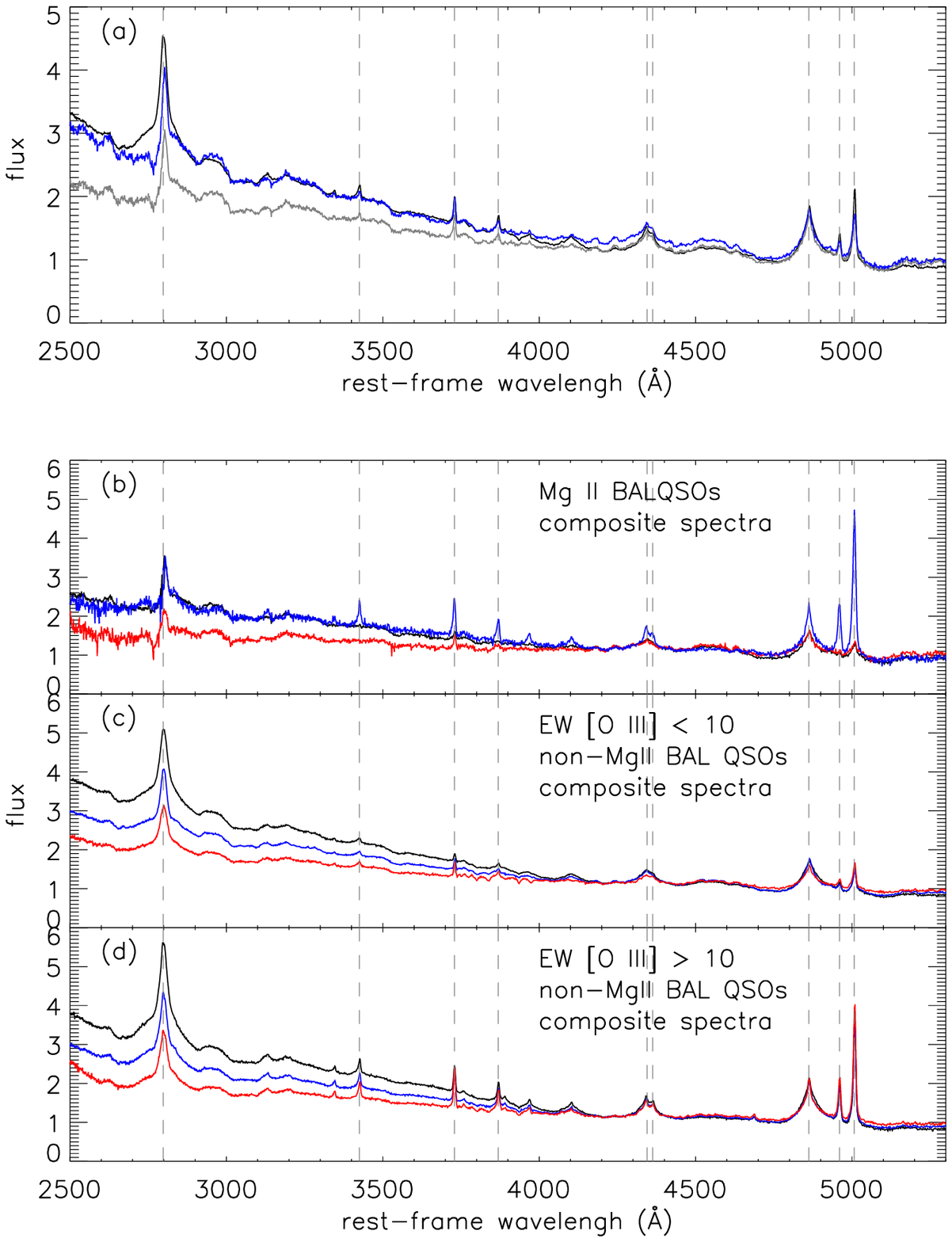} \caption{
Panel (a)--Composite spectra of the
full non-\ion{Mg}{2}  BAL quasars sample(black), \ion{Mg}{2}  BAL quasars
sample(gray) and  the blue curve is the de-reddened composite
spectra of \ion{Mg}{2}  BAL quasars;
Panel (b)--Composite spectra of the \ion{Mg}{2}  BAL
quasars with different continuum color and intensity of [\ion{O}{3}]
emission line, black for the flat and [\ion{O}{3}]-weak sources,
red means red and [\ion{O}{3}]-weak and blue curve shows the
\ion{Mg}{2}  BAL composite with flat and [\ion{O}{3}]-strong;
Panel (c) and (d)--Composite spectra of the quasars without \ion{Mg}{2}
broad absorption troughs. The three colors (black, blue and red)
means that continuum slopes are blue, flat and red. (see the text
\S3.2) }
\label{f7}
\end{figure}
\clearpage

\figurenum{8}
\begin{figure}[tbp]
\epsscale{1.0} \plotone{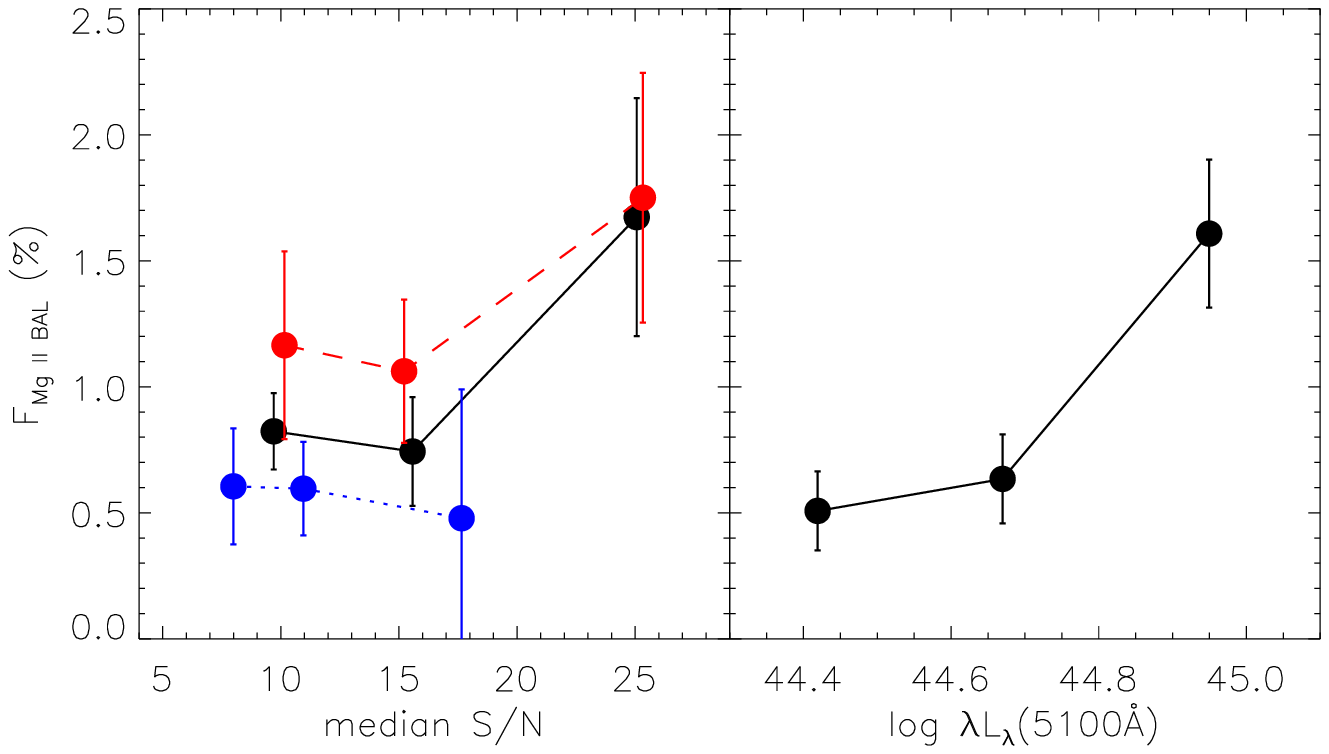} \caption{Fraction of
\ion{Mg}{2} BAL quasars in the low-$z$ sample as a function of the
median $S/N$ ratio of the SDSS spectra (left panel) and the luminosity at 5100\AA (right panel).
The median $S/N$ ratios are
calculated in the rest-frame wavelength range of $2820-3200$ \AA.
The whole sample are shown in black, and the high and low luminosity
sub-samples are shown in red and blue, respectively. The two
luminosity sub-samples are so carved that they have the same number
of quasars.
} \label{f8}
\end{figure}

\figurenum{9}
\begin{figure}[tbp]
\epsscale{1.0} \plotone{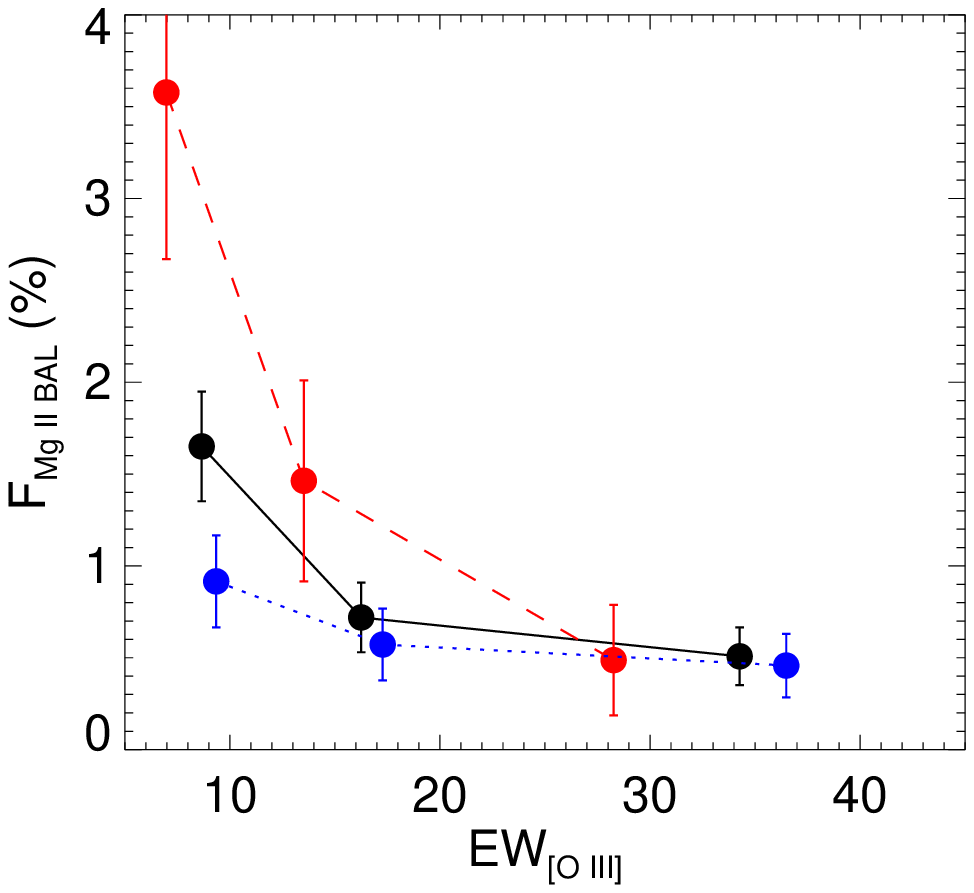} \caption{Fraction of
\ion{Mg}{2} BAL quasars observed in low-$z$ sample as a function of the
equivalent width of [\ion{O}{3}] $\lambda$5007 NEL
(black). Each point in the plot represents the fraction of
\ion{Mg}{2} BAL quasars in a bin of 2522 quasars arranged in order
of increasing $EW_{[O\;III]}$. The x-coordinate of each point
represents the median value of the bin. The "error bars" represent
1$\sigma$ limits on the fractions calculated using a binomial
distribution. We also show the connection in the two $\lambda L_{\lambda}(5100$\AA$)$
quasar bins, the red and blue lines correspond to the high and low
$\lambda L_{\lambda}(5100$\AA$)$ bins. }
 \label{f9}
\end{figure}
\clearpage

\figurenum{10}
\begin{figure}[tbp]
\epsscale{1.0} \plottwo{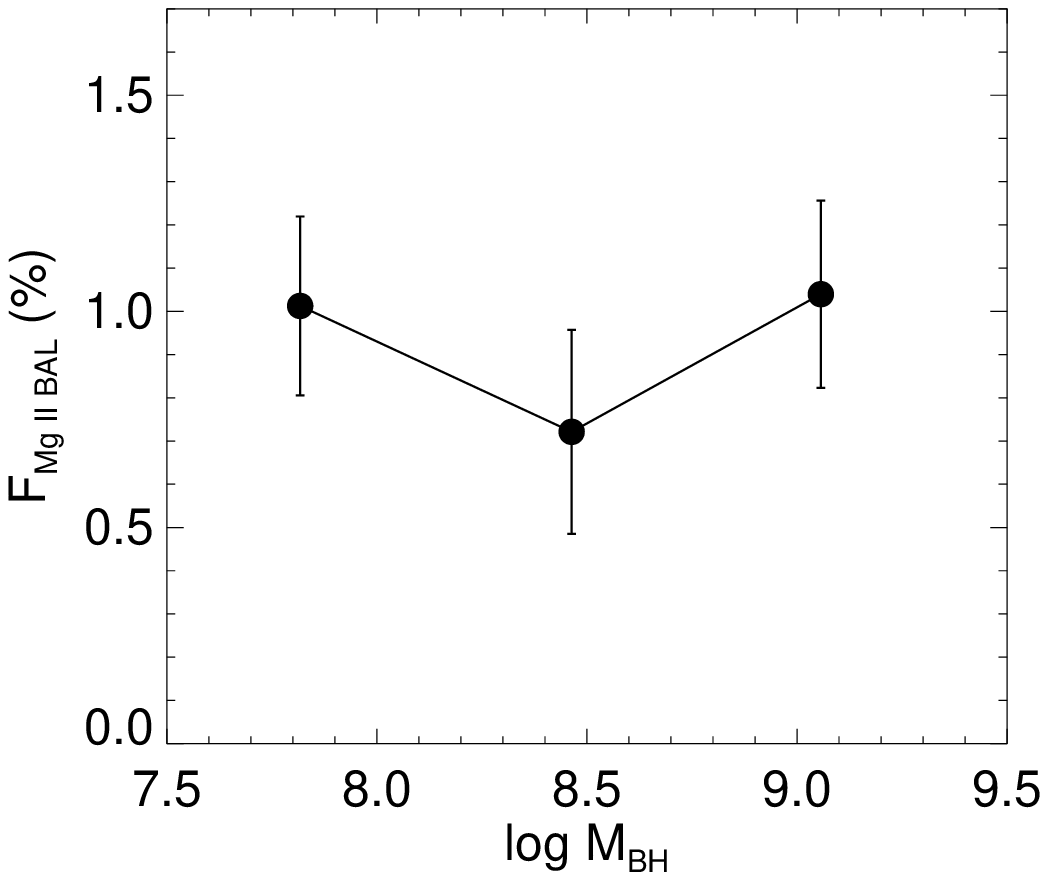}{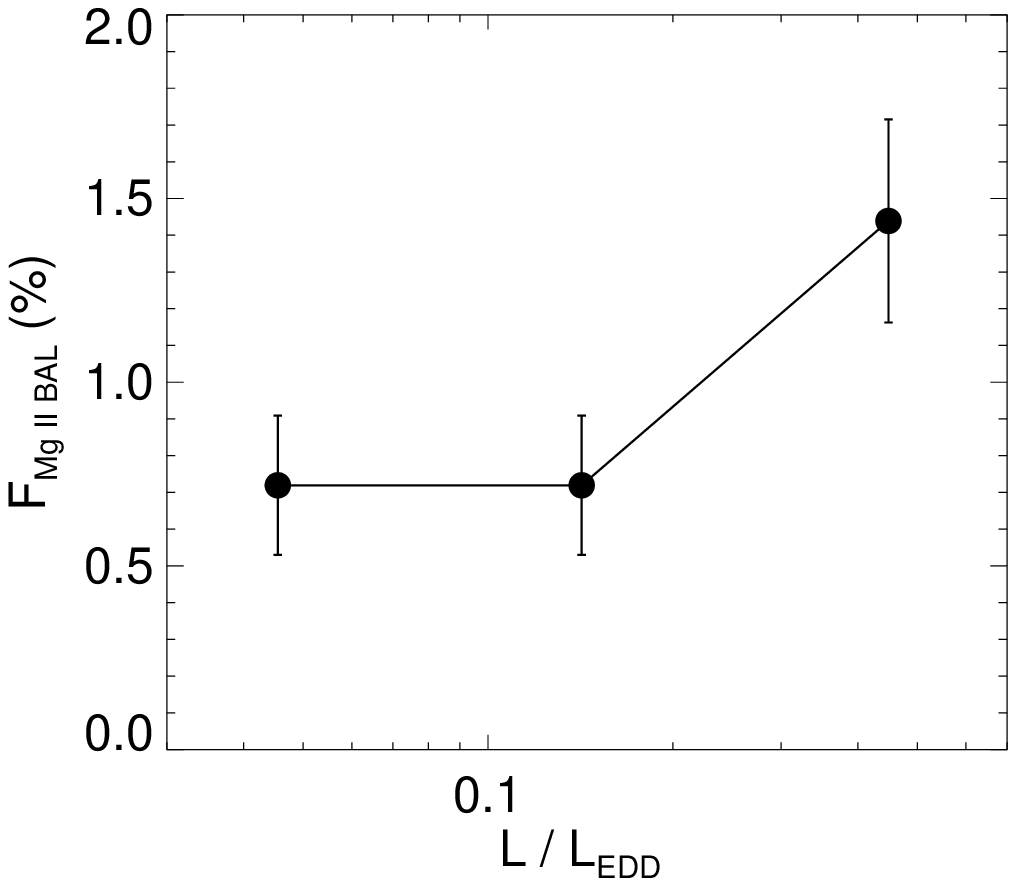}\caption{Fraction of
\ion{Mg}{2} BAL quasars in the low-$z$ sample as a function of black
hole mass (left panel) and Eddington ratio (right panel).} \label{f10}
\end{figure}
\clearpage

\figurenum{11}
\begin{figure}[tbp]
\epsscale{1.0} \plotone{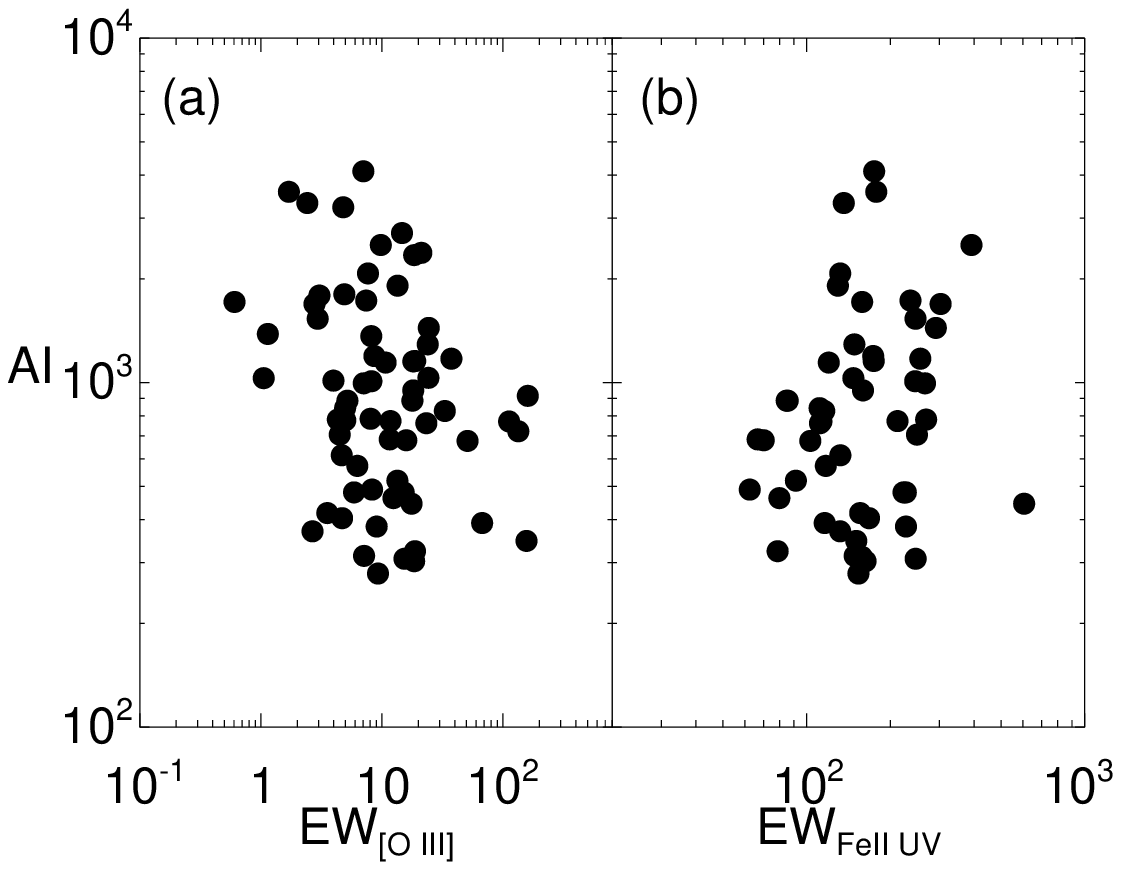}\caption{Plots of \ion{Mg}{2}
absorption index against the equivalent width of [\ion{O}{3}] (left
panel) and UV \ion{Fe}{2} (right panel) emission lines.} \label{f11}
\end{figure}
\clearpage

\figurenum{12}
\begin{figure}[tbp]
\epsscale{1.0} \plotone{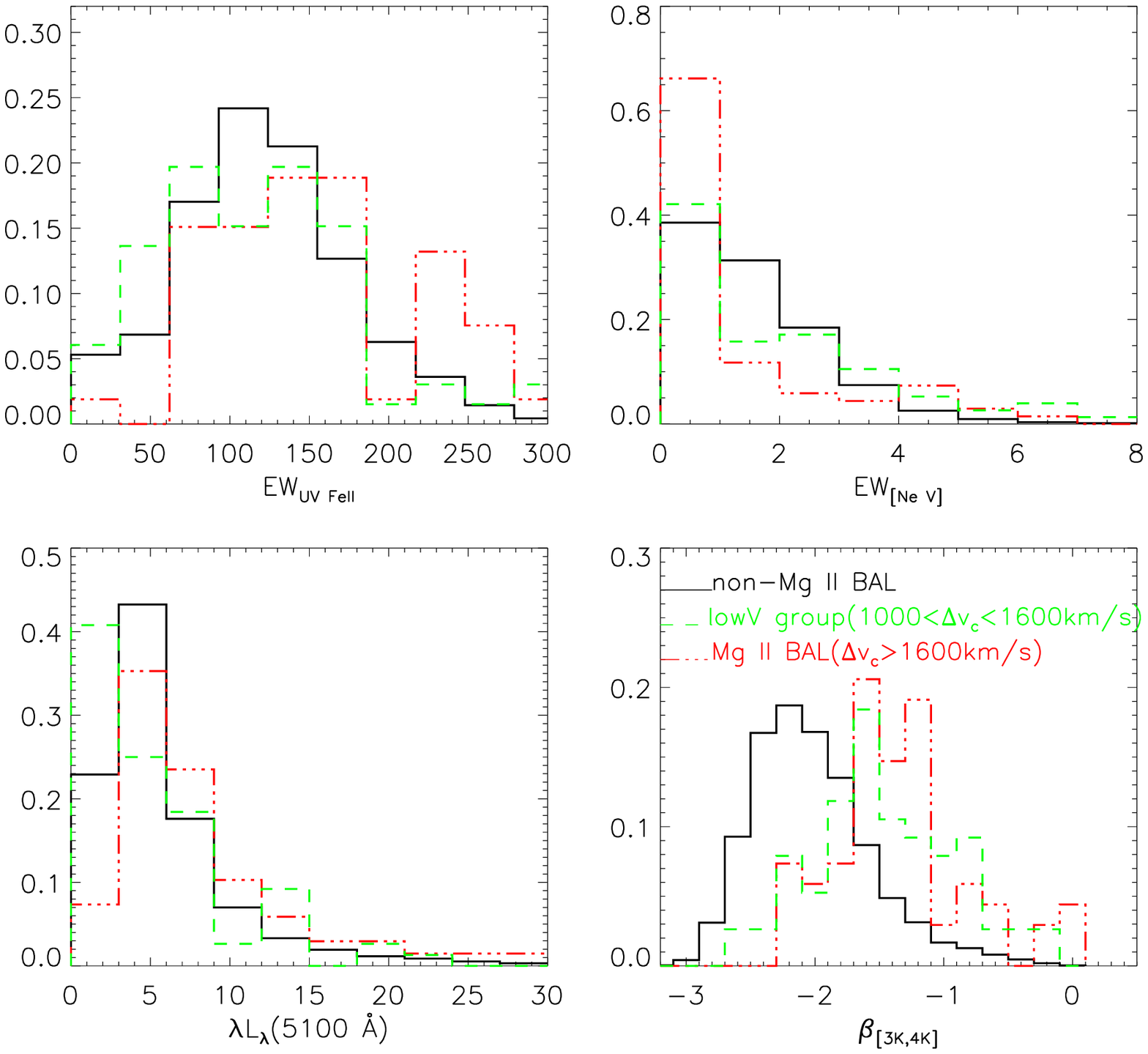}\caption{Statistical properties
of lowV group(green) and \ion{Mg}{2} BAL quasar sample (red).
The black line are non-loBAL
quasars as the comparison sample.} \label{f12}
\end{figure}
\clearpage

\figurenum{13}
\begin{figure}[tbp]
\epsscale{1.0} \plotone{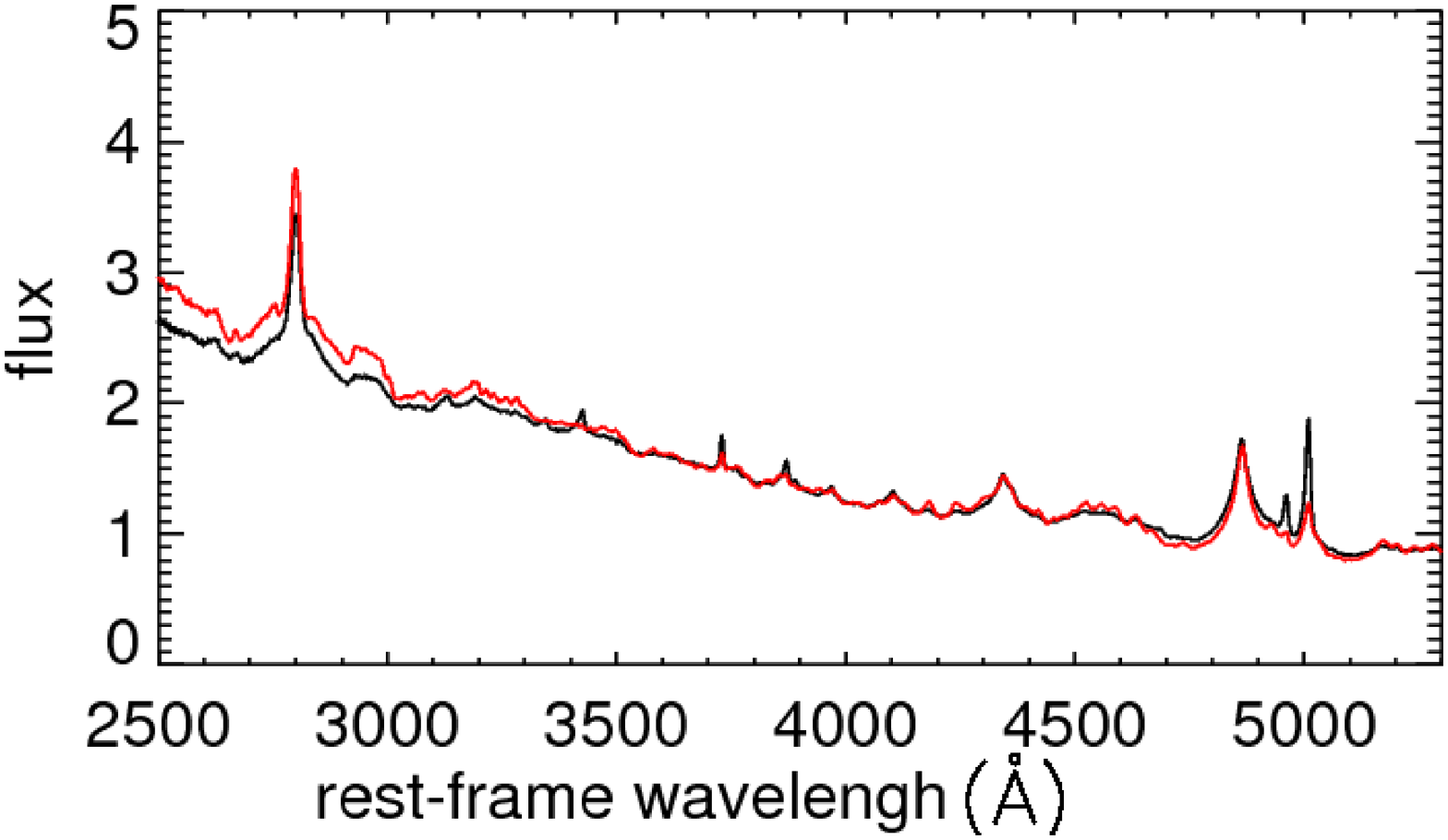}\caption{Composite spectrum of
non-loBAL quasars with/without detectable [\ion{Ne}{5}] emission.}
\label{f13}
\end{figure}
\clearpage

\end{document}